\documentclass[9pt, journal]{IEEEtran}

\usepackage{graphicx,enumerate}
\usepackage{cite,array}
\usepackage{dsfont}
\usepackage{mathrsfs}
\usepackage{amssymb}
\usepackage[cmex10]{amsmath}
\usepackage{epstopdf}

\newtheorem{theorem}{Theorem}

\newtheorem{proposition}{Proposition}
\newtheorem{definition}{Definition}
\newtheorem{remark}{Remark}

\DeclareMathOperator{\mati}{mati}
\DeclareMathOperator{\mad}{mad}

\DeclareMathOperator{\dom}{dom}
\DeclareMathOperator{\sgn}{sgn}
\DeclareMathOperator*{\esssup}{ess.\,sup}

\begin{document}

\title{Lyapunov Conditions for Input-to-State Stability of Hybrid Systems with Memory
\thanks{This work was supported by National Natural Science Foundation of China, under Grants 61374026 and 61773357.}
}

\author{Wei Ren, Junlin Xiong~\IEEEmembership{Member, IEEE}
\thanks{W.~Ren and J.~Xiong are with the Department of Automation, University of Science and Technology of China, Hefei 230026, China. \texttt{\small gtpp@ustc.mail.edu.cn}, \texttt{\small junlin.xiong@gmail.com}.}
}

\maketitle

\begin{abstract}
\boldmath
This paper studies input-to-state stability for hybrid systems with memory, which models hybrid dynamics affected by time delays. Using both Lyapunov-Razumikhin functions and Lyapunov-Krasovskii functionals, Lyapunov-based sufficient conditions are established for input-to-state stability. In addition, further extensions and relaxations are proposed for special cases, such as the stable flow/jump cases and the cases that Lyapunov functions do not decrease strictly during flow/jumps. Finally, two examples are used to illustrate the developed results.
\end{abstract}


\IEEEpeerreviewmaketitle

\section{Introduction}
\label{sec-intro}

Hybrid systems are dynamical systems with both continuous-time and discrete-time dynamics \cite{Goebel2012hybrid, Cai2007smooth}. Numerous physical and man-made systems can be modeled as hybrid systems \cite{Goebel2012hybrid}. Typical hybrid systems can be found in the literature, such as switched systems \cite{Ren2016stability} and impulsive systems \cite{Wu2016input, Liu2011input, Ren2017stability}. As a fundamental topic, stability of hybrid systems has been studied widely via Lyapunov-based approaches \cite{Goebel2012hybrid, Cai2007smooth, Cai2009characterizations, Goebel2006solutions}. On the other hand, as a ubiquitous imperfection in engineering applications, time delays have great impacts on system stability and performances. In the past few decades, a lot of attention has been given to address stability analysis for control systems with time delays; see \cite{Medvedeva2015synthesis, Zhou2016razumikhin, Heemels2010networked} and references therein.

The interplay between hybrid dynamics and time delays, which refers to hybrid systems with memory \cite{Liu2016lyapunov, Liu2014hybrid}, has attracted considerable attention, and been considered in diverse settings, such as networked control systems \cite{Heemels2010networked, Naghshtabrizi2010stability, Yuan2016delay} and chaotic systems \cite{Khadra2005impulsively, Zhu2014mean}. In particular, some typical hybrid systems with time delays have been studied in the literature; see \cite{Liu2011input, Ren2016stability} for switched time-delay systems and \cite{Chen2009input, Dashkovskiy2012stability, Wu2016input, Ren2017stability} for impulsive time-delay systems. In all these previous works, stability analysis is based on system solutions with the classic form, which means that the discrete times only lead to piecewise continuity of the system solutions but do not determine the system solutions \cite{Liu2006stability}. However, generalized solutions \cite{Goebel2012hybrid, Liu2018hybrid}, which are defined on hybrid time domains, have been applied effectively in stability analysis for hybrid systems. In \cite{Liu2018hybrid, Liu2014hybrid}, generalized solutions for hybrid systems were extended from general hybrid systems to hybrid systems with memory. In addition, the developed framework in \cite{Liu2018hybrid, Liu2014hybrid} allows us to study hybrid systems with memory via the generalized solutions.

Input-to-state stability (ISS), originally proposed in \cite{Sontag1995characterizations}, has been proven to be useful in characterizing the effects of external inputs on a control system. The ISS notation has been subsequently extended to diverse control systems, such as discrete-time systems \cite{Jiang2001input}, time-delay systems \cite{Medvedeva2015synthesis, Zhou2016razumikhin}, and hybrid systems \cite{Ren2016stability, Wu2016input, Ren2017stability, Liu2011input, Cai2009characterizations}. To study ISS of time-delay systems, the classic Lyapunov-based method is not applicable because time delays cause a violation of monotonic decrease conditions \cite{Zhou2016razumikhin, Medvedeva2015synthesis}. As a result, there are two approaches extended from the classic Lyapunov-based method. The one is based on Lyapunov-Razumikhin functions (LRFs), and the other is based on Lyapunov-Krasovskii functionals (LKFs). Such two approaches have been used widely in the previous works, such as \cite{Chen2009input, Wu2016input} using LRFs and \cite{Liu2011input, Dashkovskiy2012stability} via LKFs. To the best of our knowledge, there is no works on ISS of hybrid systems with memory via such two extended Lyapunov-based approaches, which is the topic of this paper.

In this paper, we adopt the framework in \cite{Liu2018hybrid, Liu2014hybrid}, and propose both Razumikhin-type and Krasovskii-type conditions for ISS of hybrid systems with memory. Our results extend the stability conditions for hybrid systems without memory \cite{Cai2009characterizations} along aforementioned two extended Lyapunov-based approaches. The contributions of this paper are three-fold. First, Razumikhin-type stability conditions are derived for hybrid systems with memory. In term of small gain theorem and different interpretation of Razumikhin theorem for discrete-time systems, two types of Razumikhin-type conditions are established. Moreover, the relaxed Razumikhin-type stability conditions are obtained, in which LRF is not required to decrease strictly during flow or jumps. The derived Razumikhin-type conditions extend the results in \cite{Liu2016razumikhin, Liu2016lyapunov} for pre-asymptotic stability of hybrid systems with memory. Second, Krasovskii-type stability conditions are established for ISS of hybrid systems with memory. Similar to Razumikhin-type conditions, we first obtain two types of Krasovskii-type stability conditions, and then derive some relaxed Krasovskii-type stability conditions. Our Krasovskii-type conditions extend the result in \cite[Theorem 2]{Liu2016lyapunov} for pre-asymptotic stability of hybrid systems with memory. Third, using average dwell-time like condition, we also study ISS of hybrid systems with memory in the stable flow case and the stable jump case, respectively. For such two cases, both Razumikhin-type condition and Krasovskii-type conditions are established. Finally, two numerical examples are presented to illustrate the developed results.

\section{Preliminaries}
\label{sec-preliminary}

The following notation is used throughout this paper. $\mathbb{R}:=(-\infty, +\infty)$; $\mathbb{R}_{\geq0}:=[0, +\infty)$; $\mathbb{R}_{\leq0}:=(-\infty, 0]$; $\mathbb{Z}_{\geq0}:=\{0, 1, \ldots\}$; $\mathbb{Z}_{\leq0}:=\{0, -1, \ldots\}$. $\mathbb{R}^{n}$ denotes the $n$-dimensional Euclidean space. For a given vector or matrix $A$, $A^{\top}$ denotes its transpose; $|A|$ represents the (induced) Euclidean norm. Given $(t, j), (s, k)\in\mathbb{R}^{2}$, $(t, j)\preceq(s, k)$ if $t+j\leq s+k$; $(t, j)\prec(s, k)$ if $t+j<s+k$. Given a set $\mathcal{A}\subset\mathbb{R}^{n}$ and a point $x\in\mathbb{R}^{n}$, denote by $\bar{\mathcal{A}}$ the closure of $\mathcal{A}$, and  $|x|_{\mathcal{A}}:=\inf_{y\in\mathcal{A}}|x-y|$. Given sets $\mathcal{A}, \mathcal{B}\in\mathbb{R}^{n}$, $\mathcal{A}\subset\mathcal{B}$ is relatively closed in $\mathcal{B}$ if $\mathcal{A}=\bar{\mathcal{A}}\cap\mathcal{B}$; when $\mathcal{B}$ is open, then $\mathcal{A}$ is relatively closed in $\mathcal{B}$ if and only if $\mathcal{B}\backslash\mathcal{A}$ is open. A continuous function $\alpha: \mathbb{R}_{\geq0}\to\mathbb{R}_{\geq0}$ is positive definite, denoted by $\alpha\in\mathcal{PD}$, if $\alpha(0)=0$ and $\alpha(v)>0$ for all $v>0$. A function $\alpha: \mathbb{R}_{\geq0}\to\mathbb{R}_{\geq0}$ is of class $\mathcal{K}$ if it is continuous, zero at zero and strictly increasing; $\alpha(t)$ is of class $\mathcal{K}_{\infty}$ if it is of class $\mathcal{K}$ and unbounded. A function $\beta: \mathbb{R}_{\geq0}\times\mathbb{R}_{\geq0}\to\mathbb{R}_{\geq0}$ is of class $\mathcal{KL}$ if $\beta(s, t)$ is of class $\mathcal{K}$ for each fixed $t\geq0$ and decreases to zero as $t\rightarrow\infty$ for each fixed $s\geq0$. A function $\beta: \mathbb{R}_{\geq0}\times\mathbb{R}_{\geq0}\times\mathbb{R}_{\geq0}\to\mathbb{R}_{\geq0}$ is of class $\mathcal{KLL}$ if $\beta(r, s, t)$ is of class $\mathcal{KL}$ for each fixed $t\geq0$ and of class $\mathcal{KL}$ for each fixed $s\geq0$. $\alpha^{-1}$ denotes the inverse of the function $\alpha$. For a locally Lipschitz function $V$, $V^{\circ}(x, w)$ denotes the Clarke generalized derivative of $V$ at $x$ in the direction $w$ (see also \cite{Sanfelice2014input}), i.e., $V^{\circ}(x, w)=\sup_{\zeta\in\partial V(x)}\langle\zeta, w\rangle$, where $\partial V(x)$ is the Clarke generalized gradient of $V$, which is a closed, convex, and nonempty set equal to the convex hull of all limit sequences of $\nabla V(x_{i})$ with $x_{i}\rightarrow x$ taking value away from every set of measure zero in which $V$ is nondifferentiable. A set-valued functional $\mathcal{F}: \mathcal{O}\rightrightarrows\mathbb{R}^{n}$ is outer semicontinuous at $x\in\mathcal{O}$ if for all sequence $x_{i}\rightarrow x$ and $y_{i}\rightarrow y$ with $y_{i}\in\mathcal{F}(x_{i})$, we have $y\in\mathcal{F}(x)$. A set-valued functional $\mathcal{F}: \mathcal{O}\rightrightarrows\mathbb{R}^{n}$ is locally bounded if for any compact $K\subset\mathcal{O}$, there exists a compact set $K'\subset\mathbb{R}^{n}$ such that $\mathcal{F}(K)\subset K'$.

\subsection{Hybrid Systems with Memory}

In this following, the basic knowledge on hybrid systems with memory is introduced; see \cite{Liu2014hybrid, Cai2009characterizations, Liu2016lyapunov}. We start with
the definition of hybrid time domains with memory, before introducing hybrid system with memory and its stability property.

\begin{definition}[\cite{Liu2014hybrid}]
\label{def-1}
A set $E\subseteq\mathbb{R}\times\mathbb{Z}$ is called a \emph{compact hybrid time domain with memory} if $E_{\geq0}=\bigcup^{J-1}_{j=0}([t_{j}, t_{j+1}], j)$, $E_{\leq0}=\bigcup^{K}_{k=0}([s_{k}, s_{k-1}], -k+1)$ for some finite sequence of times $s_{K}\leq\ldots\leq s_{0}=0=t_{0}\leq\ldots\leq t_{J}$. The set $E$ is called a \emph{hybrid time domain with memory} if for all $(T, J)\in E_{\geq0}$ and $(S, K)\in\mathbb{R}_{\geq0}\times\mathbb{Z}_{\geq0}$, $(E_{\geq0}\bigcap([0, T]\times\{0, \ldots, J\}))\bigcup(E_{\leq0}\bigcap([-S, 0]\times\{-K, \ldots, 0\}))$ is a compact hybrid time domain with memory. Especially, $E_{\leq0}$ is called a \emph{hybrid memory domain}.
\end{definition}

A hybrid signal is a function defined on a hybrid time domain with memory. Denote $\dom_{\geq0}x:=\dom x\bigcap(\mathbb{R}_{\geq0}\times\mathbb{Z}_{\geq0})$ and $\dom_{\leq0}(x):=\dom x\bigcap(\mathbb{R}_{\leq0}\times\mathbb{Z}_{\leq0})$. A hybrid signal $u: \dom u\mapsto\mathcal{U}$ is called a \emph{hybrid input}, if $\dom_{\leq0}(u)\equiv\{(0, 0)\}$, and $u(\cdot, j)$ is Lebesgue measurable and locally essentially bounded on $I^{j}_{u}=\{t|(t, j)\in\dom u\}$ for each $j\in\mathbb{Z}_{\geq0}$. A hybrid signal $x: \dom x\mapsto\mathbb{R}^{n}$ is called a \emph{hybrid arc with memory}, if $x(\cdot, j)$ is locally absolutely continuous on $I^{j}_{x}=\{t|(t, j)\in\dom x\}$ for each $j\in\mathbb{Z}$. A \emph{hybrid memory arc} is a hybrid arc with memory whose time domain is hybrid memory domain. The set of all hybrid memory arcs is denoted by $\mathcal{M}$.

Given a hybrid time domain with memory $E$, define $\sup_{t\geq0}E:=\{t\in\mathbb{R}_{\geq0}| \exists j\in\mathbb{Z}_{\geq0}\text{ such that }(t, j)\in E_{\geq0}\}$ and $\inf_{t\leq0}E:=\{t\in\mathbb{R}_{\leq0}| \exists j\in\mathbb{Z}_{\leq0}\text{ such that }(t, j)\in E_{\leq0}\}$. Similarly, $\sup_{j\geq0}E$ and $\inf_{j\geq0}E$ can be defined. Denote  $\mathfrak{L}_{\geq0}(E):=\sup_{t\geq0}E+\sup_{j\geq0}E$ and $\mathfrak{L}_{\leq0}(E):=\sup_{t\leq0}E+\sup_{j\leq0}E$. Given $\Delta\in\mathbb{R}_{\geq0}$, $\mathcal{M}^{\Delta}$ denotes the set of the hybrid memory arcs $\varphi$ satisfying $-\Delta-1\leq\mathfrak{L}_{\leq0}(\dom\varphi)\leq-\Delta$. Given a hybrid arc with memory $x$, the operator $\mathcal{A}^{\Delta}_{[\cdot, \cdot]}x: \dom_{\geq0}x\mapsto\mathcal{M}^{\Delta}$ is defined by $\mathcal{A}^{\Delta}_{[t, j]}x(s, k)=x(t+s, j+k)$ for all $(s, k)\in\dom\mathcal{A}^{\Delta}_{[t, j]}x$, where $(t, j)\in\dom_{\geq0}x$, $\dom\mathcal{A}^{\Delta}_{[t, j]}x:=\{(s, k)\in\mathbb{R}_{\leq0}\times\mathbb{Z}_{\leq0} | (t+s, j+k)\in\dom x, s+k\geq-\Delta_{\inf}\}$, and $\Delta_{\inf}:=\inf\{\delta\geq\Delta|\exists (t+s, j+k)\in\dom x \text{ such that } s+k=-\delta\}$. For the sake of simplicity, $\mathcal{A}^{\Delta}x:=\mathcal{A}^{\Delta}_{[\cdot, \cdot]}x$ if the time argument is omitted.

Consider the following hybrid system with memory of size $\Delta>0$, denoted by $\mathcal{H}^{\Delta}_{\mathcal{M}}:=(\mathcal{F}, \mathcal{G}, \mathcal{C}, \mathcal{D}, \mathbb{R}^{n}, \mathcal{U})$,
\begin{align}
\label{eqn-1}
\begin{cases}
\dot{x}\in \mathcal{F}(\mathcal{A}^{\Delta}x, u), \quad &\forall (\mathcal{A}^{\Delta}x, u)\in\mathcal{C}, \\
x^{+}\in \mathcal{G}(\mathcal{A}^{\Delta}x, u), \quad &\forall (\mathcal{A}^{\Delta}x, u)\in\mathcal{D},
\end{cases}
\end{align}
where $x\in\mathbb{R}^{n}$ is the system state, $u\in\mathcal{U}\subset\mathbb{R}^{m}$ is the external input, $\mathcal{C}\subseteq\mathcal{M}^{\Delta}\times\mathcal{U}$ is the flow set and $\mathcal{D}\subseteq\mathcal{M}^{\Delta}\times\mathcal{U}$ is the jump set. The following regularity assumptions on the data $(\mathcal{F}, \mathcal{G}, \mathcal{C}, \mathcal{D}, \mathbb{R}^{n}, \mathcal{U})$ are imposed in this paper; see also \cite{Liu2014hybrid}.
\begin{enumerate}[({SA}1)]
  \item $\mathcal{U}\subset\mathbb{R}^{m}$ is closed, and $\mathcal{C}\cap(\mathcal{M}^{\Delta}\times\mathcal{U})$ and $\mathcal{D}\cap(\mathcal{M}^{\Delta}\times\mathcal{U})$ are relatively closed in $\mathcal{M}^{\Delta}\times\mathcal{U}$;
  \item $\mathcal{F}: \mathcal{C}\rightrightarrows\mathbb{R}^{n}$ is outer semicontinuous and locally bounded relative to the set $\mathcal{C}$, and $\mathcal{F}(\varphi, u)$ is nonempty and convex for each $(\varphi, u)\in\mathcal{C}$.
  \item $\mathcal{G}: \mathcal{D}\rightrightarrows\mathbb{R}^{n}$ is outer semicontinuous and locally bounded relative to the set $\mathcal{D}$, and $\mathcal{G}(\varphi, u)$ is nonempty for each $(\varphi, u)\in\mathcal{D}$.
\end{enumerate}

\begin{remark}
\label{rmk-0}
The regularity assumptions are similar to those required for hybrid systems without memory in \cite{Sanfelice2014input, Cai2009characterizations}. These assumptions enable the developments in this work, and assure some properties for $\mathcal{H}^{\Delta}_{\mathcal{M}}$, such as well-posedness of hybrid systems and existence of generalized solutions; see also \cite{Liu2014hybrid, Liu2018hybrid}.
\hfill $\square$
\end{remark}

\begin{definition}
\label{def-3}
A hybrid arc with memory $x$ and a hybrid input $u$ are a \emph{solution pair} $(x, u)$ to $\mathcal{H}^{\Delta}_{\mathcal{M}}$, if
\begin{enumerate}[{(i)}]
  \item $\dom_{\geq0}x=\dom u$ and $(\mathcal{A}^{\Delta}_{[0, 0]}x, u(0, 0))\in\mathcal{C}\cup\mathcal{D}$;
  \item for all $j\in\mathbb{Z}_{\geq0}$ and almost all $t$ such that $(t, j)\in\dom_{\geq0}x$,
  \begin{equation}
  \label{eqn-2}
  (\mathcal{A}^{\Delta}_{[t, j]}x, u(t, j))\in\mathcal{C}, \quad   \dot{x}(t, j)\in\mathcal{F}(\mathcal{A}^{\Delta}_{[t, j]}x, u(t, j));
  \end{equation}
  \item for all $(t, j)\in\dom_{\geq0}x$ such that $(t, j+1)\in\dom_{\geq0}x$,
  \begin{equation}
  \label{eqn-3}
  (\mathcal{A}^{\Delta}_{[t, j]}x, u(t, j))\in\mathcal{D}, \quad  x(t, j+1)\in\mathcal{G}(\mathcal{A}^{\Delta}_{[t, j]}x, u(t, j)).
  \end{equation}
\end{enumerate}
\end{definition}

Given any hybrid signal $z$, denote $\sharp z:=\sup_{(t, j)\in\dom z}t+j$. For any hybrid input $u: \dom u\mapsto\mathcal{U}$, and let $(t_{1}, j_{1}), (t_{2}, j_{2})\in\dom u$ and $(t_{1}, j_{1})\preceq(t_{2}, j_{2})$, define (see also \cite{Cai2009characterizations})
\begin{align*}
\|u\|_{[(t_{1}, j_{1}), (t_{2}, j_{2})]}&:=\max\left\{\sup\limits_{(t, j)\in\Gamma(u), (t_{1}, j_{1})\preceq(t, j)\preceq(t_{2}, j_{2})}|u(t, j)|,\right.\\
&\quad \left.\esssup\limits_{(t, j)\in\dom u\setminus\Gamma(u), (t_{1}, j_{1})\preceq(t, j)\preceq(t_{2}, j_{2})}|u(t, j)|\right\},
\end{align*}
where $\Gamma(u):=\{(t, j)\in\dom u|(t, j+1)\in\dom u\}$. Denote $\|u\|_{(t_{2}, j_{2})}:=\|u\|_{[(0, 0), (t_{2}, j_{2})]}$ and $\|u\|_{\sharp}:=\lim_{t_{2}+j_{2}\rightarrow\sharp u}\|u\|_{(t_{2}, j_{2})}$. In particular, $\|u\|_{\infty}:=\|u\|_{\sharp}$ as $\sharp u=+\infty$.

A solution pair $(x, u)$ to $\mathcal{H}^{\Delta}_{\mathcal{M}}$ is nontrivial if $\dom_{\geq0}x$ has at least two elements; it is complete if $\dom_{\geq0}x$ is unbounded; it is maximal if it cannot be extended; and it is bounded if there exists a compact set $\mathcal{O}\in\mathbb{R}^{n}$ such that $x(t, j)\in\mathcal{O}$ for all $(t, j)\in\dom x$. $\mathfrak{S}^{\Delta}_{u}(\varphi)$ denotes the set of all the maximal solution pairs $(x, u)$ to $\mathcal{H}^{\Delta}_{\mathcal{M}}$ with the initial $\varphi\in\mathcal{M}^{\Delta}$ and finite $\|u\|_{\sharp}$. Assume that the system $\mathcal{H}^{\Delta}_{\mathcal{M}}$ is forward complete, that is, for all $\varphi\in\mathcal{M}^{\Delta}$, the solution pair $(x, u)\in\mathfrak{S}^{\Delta}_{u}(\varphi)$ is complete. The objective of this paper is to achieve Lyapunov-based conditions to guarantee input-to-state stability of hybrid systems with memory.

\begin{definition}
\label{def-4}
Consider the hybrid system with memory $\mathcal{H}^{\Delta}_{\mathcal{M}}$, a closed set $\mathcal{W}\subset\mathbb{R}^{n}$ is \emph{input-to-state stable (ISS)}, if there exist $\beta\in\mathcal{KLL}, \gamma\in\mathcal{K}$ such that for all $(t, j)\in\dom_{\geq0}x$ and all $\mathcal{A}^{\Delta}_{[0, 0]}x\in\mathcal{M}^{\Delta}$, each solution pair $(x, u)\in\mathfrak{S}^{\Delta}_{u}(\varphi)$ satisfies
\begin{equation}
\label{eqn-4}
|x(t, j)|_{\mathcal{W}}\leq\max\{\beta(\|\mathcal{A}^{\Delta}_{[0, 0]}x\|_{\mathcal{W}}, t, j), \gamma(\|u\|_{(t, j)})\},
\end{equation}
where $\|\varphi\|_{\mathcal{W}}:=\sup_{s+k\in[-\Delta-1, 0]}|\varphi(s, k)|_{\mathcal{W}}$ for $\varphi\in\mathcal{M}^{\Delta}$.
\end{definition}

\section{ISS with Lyapunov-Razumikhin Functions}
\label{sec-razumikhin}

In this section, we show different types of Razumikhin-type stability conditions for the system \eqref{eqn-1}. We start with standard Razumikhin-type conditions, before presenting some useful extensions like for the system \eqref{eqn-1} satisfying persistent flow.

\subsection{Razumikhin-type Stability Conditions}

According to small gain theorem, the following two typical Razumikhin-type conditions are established for ISS of $\mathcal{H}^{\Delta}_{\mathcal{M}}$.

\begin{theorem}
\label{thm-1}
Consider the hybrid system with memory $\mathcal{H}^{\Delta}_{\mathcal{M}}$. Let $\Delta<\infty$ and $\mathcal{W}\subset\mathbb{R}^{n}$ be closed. If there exist a locally Lipschitz function $V: \mathbb{R}^{n}\rightarrow\mathbb{R}_{\geq0}, \alpha_{1}, \alpha_{2}\in\mathcal{K}_{\infty}, \alpha_{3}, \rho\in\mathcal{PD}$, and continuous nondecreasing functions $\gamma_{1}, \gamma_{2}: \mathbb{R}_{\geq0}\rightarrow\mathbb{R}_{\geq0}$ such that
\begin{enumerate}[({A}.1)]
  \item  for all $\varphi\in\mathcal{M}^{\Delta}$, $\alpha_{1}(|\varphi(0, 0)|_{\mathcal{W}})\leq V(\varphi(0, 0))\leq\alpha_{2}(|\varphi(0, 0)|_{\mathcal{W}})$;
  \item  for all $(\varphi, u)\in\mathcal{C}$ and all $f\in\mathcal{F}(\varphi, u)$, $V(\varphi(0, 0))\geq\max\{\gamma_{1}(\bar{V}(\varphi)), \gamma_{2}(|u|)\}$ implies that $V^{\circ}(\varphi(0, 0), f)\leq-\alpha_{3}(V(\varphi(0, 0)))$;
  \item  for all $(\varphi, u)\in\mathcal{D}$ and all $g\in\mathcal{G}(\varphi, u)$, $V(g)\leq\rho(\bar{V}(\varphi))$;
  \item $\rho$ and $\gamma_{1}$ satisfy the small gain condition, that is, $\rho(v)<v$ and $\gamma_{1}(v)<v$ for all $v>0$,
\end{enumerate}
where $\bar{V}(\varphi):=\sup_{(s, k)\in\dom\varphi, s+k\geq-\Delta-1}V(\varphi(s, k))$, then the set $\mathcal{W}$ is ISS for $\mathcal{H}^{\Delta}_{\mathcal{M}}$.
\end{theorem}

\begin{IEEEproof}
According to (A.2), we divide the proof into the following two cases: the case that $V(\varphi(0, 0))\geq\max\{\gamma_{1}(\bar{V}(\varphi)), \gamma_{2}(|u|)\}$ and the case that $V(\varphi(0, 0))\leq\max\{\gamma_{1}(\bar{V}(\varphi)), \gamma_{2}(|u|)\}$.

For the first case, we prove that there exists $\beta\in\mathcal{KLL}$ such that
\begin{equation}
\label{eqn-5}
V(x(t, j))\leq\beta(\bar{V}(\mathcal{A}^{\Delta}_{[0, 0]}x), t, j), \quad \forall (t, j)\in\dom_{\geq0}x.
\end{equation}

To this end, We first prove that $\bar{V}(\mathcal{A}^{\Delta}_{[t, j]}x)$ is non-increasing for $(t, j)\in\dom x$. If $(t, j)\in\dom x$ and $(t, j+1)\in\dom x$, we get from (A.3)-(A.4) that $V(x(t, j+1))\leq\rho(\bar{V}(\mathcal{A}^{\Delta}_{[t, j]}x))<\bar{V}(\mathcal{A}^{\Delta}_{[t, j]}x)$, which implies that $\bar{V}(\mathcal{A}^{\Delta}_{[t, j+1]}x)\leq\bar{V}(\mathcal{A}^{\Delta}_{[t, j]}x)$. If $(t+\Delta t, j)\in\dom x$ for all $\Delta t\in[0, h]$ with some small $h>0$, then there are two cases: (i) $V (x(t, j))=\bar{V}(\mathcal{A}^{\Delta}_{[t, j]}x)$, and (ii) $V(x(t, j))<\bar{V}(\mathcal{A}^{\Delta}_{[t, j]}x)$. For the case (i), we prove that $V(x(t+\Delta t, j))\leq V(x(t, j))$ for all $\Delta t\in[0, h]$. To this end, we show that for any given $\varepsilon>0$, $V(x(t+\Delta t, j))<V(x(t, j))+\varepsilon$ for all $\Delta t\in[0, h]$. If not, then define $h_{1}:=\inf\{\Delta t\in[0, h]|V (x(t +\Delta t, j))\geq V (x(t, j))+\varepsilon\}$. In the sequel, we have that
\begin{align*}
V (x(t+h_{1}, j))&=V(x(t, j)) +\varepsilon,  \\
V (x(t+\Delta t, j))&<V(x(t, j)) +\varepsilon, \quad \forall \Delta t\in[0, h_{1}).
\end{align*}
It follows from (A.2) that for almost all $\Delta t\in[0, h_{1}]$,
\begin{equation*}
\frac{dV(x(t+\Delta t, j))}{dt}\leq-\alpha_{3}(V(x(t+\Delta t, j))),
\end{equation*}
which implies that $V(x(t+h_{1}, j))<V(x(t+\Delta t, j))<V(x(t, j))+\varepsilon$, which is a contradiction. Hence, $V(x(t+\Delta t, j))\leq V(x(t, j))$ for all $\Delta t\in[0, h]$, which implies that $\bar{V}(\mathcal{A}^{\Delta}_{[t+\Delta t, j]}x)\leq\bar{V}(\mathcal{A}^{\Delta}_{[t, j]}x)$. For the case (ii), it follows from the continuity of $x(t+\Delta t, j)$ that $\bar{V}(\mathcal{A}^{\Delta}_{[t+\Delta t, j]}x)\leq\bar{V}(\mathcal{A}^{\Delta}_{[t, j]}x)$ for all $\Delta t\in[0, h]$. Based to above analysis, $\bar{V}(\mathcal{A}^{\Delta}_{[t,j]}x)$ is non-increasing for $(t, j)\in\dom x$. Define $\varrho:=\|\mathcal{A}^{\Delta}_{[0, 0]}x\|_{\mathcal{W}}$, and it follows from (A.1) that for all $(t, j)\in\dom_{\geq0}x$,
\begin{align}
\label{eqn-6}
V(x(t, j))&\leq\bar{V}(\mathcal{A}^{\Delta}_{[t,j]}x)\nonumber\\
&\leq\bar{V}(\mathcal{A}^{\Delta}_{[0, 0]}x)\leq\alpha_{2}(\varrho).
\end{align}

Second, we prove that for all $\varepsilon\in[0, \alpha_{2}(\varrho)]$, there exists certain $T_{1}>0$ such that $V (x(t, j))\leq\varepsilon$ for all $(t, j)\in\dom_{\geq0}x$ with $t+j\geq T_{1}$. Define $\delta_{1}:=\min_{v\in[\varepsilon, \alpha_{2}(\varrho)]}\{\alpha_{3}(v), v-\rho(v)\}$, $\delta_{2}:=\min_{v\in[\varepsilon, \alpha_{2}(\varrho)]}\{v-\rho(v)\}$, and $N:=\inf\{k\in\mathbb{Z}_{>0}| \varepsilon+k\delta_{2}\geq\alpha_{2}(\varrho)\}$.

Claim that there exists certain $(\bar{t}, \bar{j})\in\dom x$ and $\bar{t}+\bar{j}\geq0$ such that $V (x(\bar{t}, \bar{j}))\leq\varepsilon+(N-1)\delta_{2}$. If not, then $V(x(t, j))>\varepsilon+(N-1)\delta_{2}$ for all $(t, j)\in\dom x$ and $t+j\geq0$. It follows from (A.2) that $dV(x(t, j))/dt\leq-\alpha_{3}(V(x(t, j)))\leq-\delta_{1}$. If $(t, j)\in\dom x$ and $(t, j+1)\in\dom x$, we have from (A.3) that $V(x(t, j+1))-V(x(t, j))\leq\rho(\bar{V}(\mathcal{A}^{\Delta}_{[t, j]}x))-V (x(t, j))\leq-\delta_{1}$. As a result, $V(x(t, j))\leq V(x(0, 0))-\delta_{1}(t+j)$ for all $(t, j)\in\dom x$ and $t+j\geq0$. This leads to a contradiction if $t+j$ is sufficiently large.

Claim that $V (x(t, j))\leq\varepsilon+(N-1)\delta_{2}$ for all $(t, j)\in\dom x$ and $(t, j)\succ(\bar{t}, \bar{j})$. If not, then there exists $(t', j')\in\dom x$ such that $t'+j'=\inf\{t+j\geq \bar{t}+\bar{j}|V(x(t, j))>\varepsilon+(N-1)\delta_{2}\}$. If $V(x(t', j'))\geq\varepsilon+(N-1)\delta_{2}$ due to the jump, then $(t', j'-1)\in\dom x$ and $V(x(t', j'-1))\leq\varepsilon+(N-1)\delta_{2}$. As a result, it follows from (A.3)-(A.4) that $V(x(t', j'))\leq\rho(\bar{V}(\mathcal{A}^{\Delta}_{[t', j'-1]}x))<\bar{V}(\mathcal{A}^{\Delta}_{[t', j'-1]}x)\leq\varepsilon+(N-1)\delta_{2}$, which is a contradiction. If $V (x(t', j')\geq\varepsilon+(N-1)\delta_{2}$ due to the flow, then it follows from (A.2) that $dV(x(t', j'))/dt\leq-\alpha_{3}(V(x(t', j'))<0$, which implies that $V(x(t', j'))<V(x(t'-\Delta t, j'))<\varepsilon+(N-1)\delta_{2}$. This contradicts with the definition of $(t', j')$.

Combining such two claims leads to $\bar{V}(\mathcal{A}^{\Delta}_{[t, j]}x)\leq\varepsilon+(N-1)\delta_{2}$, for all $t+j\geq T_{1}+\Delta+1$. Repeating the same argument above, we can inductively show that, there exists $T>T_{1}$ such that
\begin{align}
\label{eqn-7}
V(x(t, j))\leq\bar{V}(\mathcal{A}^{\Delta}_{[t, j]}x)\leq\varepsilon, \quad \forall t+j\geq T.
 \end{align}
As a result, it follows from \eqref{eqn-6}-\eqref{eqn-7} that there exists $\beta\in\mathcal{KLL}$ such that \eqref{eqn-5} holds for all $(t, j)\in\dom_{\geq0}x$.

Combining \eqref{eqn-5} with the second case yields that for all $(t, j)\in\dom_{\geq0}x$,
\begin{align}
\label{eqn-8}
V(x(t, j))&\leq\max\{\beta(\bar{V}_{[0, 0]}(x), t, j), \gamma_{1}(\|\bar{V}\|_{[0, L)}),  \nonumber\\
&\quad \gamma_{2}(\|u\|_{(t, j)})\}, \\
\label{eqn-9}
\bar{V}_{[t, j]}(x)&\leq\max\{\bar{V}_{[0, 0]}(x)\cdot\phi(t+j), \|V\|_{[0, L)}\},
\end{align}
where $\bar{V}_{[t, j]}(x):=\bar{V}(\mathcal{A}^{\Delta}_{[t, j]}x)$, $L:=\mathfrak{L}_{\geq0}(\dom x)$, $\|\bar{V}\|_{[0, L)}:=\sup_{t+j\in[0, L)}\bar{V}_{[t, j]}(x)$, $\|V\|_{[0, L)}:=\sup_{t+j\in[0, L)}V(x(t, j))$, and $\phi(v):=0.5(1-\sgn(v-\Delta-1))$; $\sgn(v)=1$ if $v\geq0$ and $\sgn(v)=-1$ if $v<0$.

Taking the sup norm on both sides of \eqref{eqn-8}-\eqref{eqn-9}, substituting \eqref{eqn-9} into \eqref{eqn-8} and using (A.4), we obtain that for all $(t, j)\in\dom_{\geq0}x$,
\begin{align}
\label{eqn-10}
\|\bar{V}\|_{[0, L)}&\leq\max\{\beta(\bar{V}_{[0, 0]}(x), 0, 0), \gamma_{2}(\|u\|_{\sharp})\}.
\end{align}

Given $\varepsilon, \eta_{1}, \eta_{2}>0$, define $\delta:=\max\{\beta(\alpha_{2}(\eta_{1}), 0, 0), \gamma_{2}(\eta_{2})\}$. Choose $T_{1}, J_{1}>0$ such that $\beta(\delta, T_{1}, J_{1})\leq\alpha_{1}(\varepsilon)$. Furthermore, let $T_{2}+J_{2}>\Delta+1$, it follows from \eqref{eqn-8}-\eqref{eqn-9} that
\begin{align}
\label{eqn-11}
&\|V\|_{[T_{1}+T_{2}+J_{1}+J_{2}, L)}\leq\|V\|_{[T_{2}+J_{2}, L)}  \nonumber \\
&\qquad \qquad\qquad\qquad\leq\max\{\alpha_{1}(\varepsilon), \gamma_{1}(\|\bar{V}\|_{[0, L)}), \gamma_{2}(\|u\|_{\sharp})\}.
\end{align}
Because $\gamma_{1}(v)<v$ for all $v>0$, there exists $n(\delta, \varepsilon)\in\mathbb{Z}_{>0}$ such that $\gamma^{n}_{1}(\delta)\leq\max\{\alpha_{1}(\varepsilon), \gamma_{2}(\|u\|_{\sharp})\}$. It obtains from \eqref{eqn-11} that
\begin{align}
\label{eqn-12}
\|V\|_{[n(T_{1}+T_{2}+J_{1}+J_{2}), L)}&\leq\max\{\alpha_{1}(\varepsilon), \gamma_{2}(\|u\|_{\sharp})\},
\end{align}
which implies that $|x(t, j)|_{\mathcal{W}}\leq\max\{\varepsilon, \alpha^{-1}_{1}(\gamma_{2}(\|u\|_{\sharp}))\}$ for all $(t, j)\in\dom_{\geq0}x$ with $t+j\geq n(T_{1}+T_{2}+J_{1}+J_{2})$. As a result, according to Proposition 2.12 and Theorem 3.1 in \cite{Cai2009characterizations}, it follows from \eqref{eqn-6}, \eqref{eqn-10} and \eqref{eqn-12} that the set $\mathcal{W}$ is ISS for $\mathcal{H}^{\Delta}_{\mathcal{M}}$.
\end{IEEEproof}

\begin{remark}
\label{rmk-1}
The proof techniques for Theorem \ref{thm-1} follows \cite{Teel1998connections}, which is on continuous time-delay systems. The inequality \eqref{eqn-10} implies that $\mathcal{W}$ is globally pre-stable for $\mathcal{H}^{\Delta}_{\mathcal{M}}$, whereas \eqref{eqn-12} implies that $\mathcal{H}^{\Delta}_{\mathcal{M}}$ has the asymptotic gain property; see \cite{Sontag1996new, Cai2009characterizations} for further details. As a result, Theorem \ref{thm-1} is an extension of the classic stability results; such as \cite{Teel1998connections} on functional differential equations and \cite{Liu2016razumikhin, Liu2016lyapunov} on asymptotic stability of hybrid systems with memory.
\hfill $\square$
\end{remark}

\begin{theorem}
\label{thm-2}
Consider the system $\mathcal{H}^{\Delta}_{\mathcal{M}}$. Let $\Delta<\infty$ and $\mathcal{W}\subset\mathbb{R}^{n}$ be closed. If there exist a locally Lipschitz function $V: \mathbb{R}^{n}\rightarrow\mathbb{R}_{\geq0}, \alpha_{1}, \alpha_{2}\in\mathcal{K}_{\infty}, \rho\in\mathcal{PD}$, continuous nondecreasing functions $\gamma_{1}, \gamma_{2}: \mathbb{R}_{\geq0}\rightarrow\mathbb{R}_{\geq0}$ such that (A.1) holds, $\gamma_{1}$ satisfies the small gain condition and
\begin{enumerate}[(B.1)]
  \item for all $(\varphi, u)\in\mathcal{C}$ and all $f\in\mathcal{F}(\varphi, u)$, $V(\varphi(0, 0))\geq\max\{\gamma_{1}(\bar{V}(\varphi)), \gamma_{2}(|u|)\}$ implies that $V^{\circ}(\varphi(0, 0), f)\leq-\rho(|\varphi(0, 0)|_{\mathcal{W}})$;
  \item for all $(\varphi, u)\in\mathcal{D}$ and all $g\in\mathcal{G}(\varphi, u)$, $V(\varphi(0, 0))\geq\max\{\gamma_{1}(\bar{V}(\varphi)), \gamma_{2}(|u|)\}$ implies that $V(g)-V(\varphi(0, 0))\leq-\rho(|\varphi(0, 0)|_{\mathcal{W}})$,
\end{enumerate}
then the set $\mathcal{W}$ is ISS for $\mathcal{H}^{\Delta}_{\mathcal{M}}$.
\end{theorem}

\begin{IEEEproof}
Similar to the proof of Theorem \ref{thm-1}, we consider two cases. For the first case, 
following the similar line as in the proof of Theorem \ref{thm-1}, we have that there exists $\beta\in\mathcal{KLL}$ such that
\begin{equation}
\label{eqn-13}
V(x(t, j))\leq\beta(\bar{V}_{[0, 0]}(x), t, j),  \quad \forall (t, j)\in\dom_{\geq0}x.
\end{equation}
It follows from \eqref{eqn-13} and the second case that for all $(t, j)\in\dom_{\geq0}x$,
\begin{align*}
\bar{V}_{[t, j]}(x)&\leq\max\{\bar{V}_{[0, 0]}(x)\cdot\phi(t+j), \|V\|_{[0, L)}\}, \\
V(x(t, j))&\leq\max\{\beta(\bar{V}_{[0, 0]}(x), t, j), \gamma_{1}(\|\bar{V}\|_{[0, L)}), \\ 
&\quad \gamma_{2}(\|u\|_{(t, j)})\}.
\end{align*}
The following is along the same fashion as the proof of Theorem \ref{thm-1}. As a result, the set $\mathcal{W}$ is ISS for $\mathcal{H}^{\Delta}_{\mathcal{M}}$.
\end{IEEEproof}

\begin{remark}
\label{rmk-2}
Razumikhin-type conditions in Theorems \ref{thm-1}-\ref{thm-2} are equivalent. The difference between Theorem \ref{thm-1} and Theorem \ref{thm-2} lies in (A.3) and (B.2). Compared to (A.3), (B.2) provides a direct interpretation of Razumikhin theorem for discrete-time systems; see also \cite{Liu2016razumikhin}.
\hfill $\square$
\end{remark}

\subsection{Extensions of Razumikhin-type Stability Conditions}

In this subsection, some extensions of Razumikhin-type conditions are presented, which do not require LRFs to decrease strictly during the flow or at the jumps. First, the following two relaxed results provides sufficient conditions for $\mathcal{H}^{\Delta}_{\mathcal{M}}$ satisfying persistent flow and persistent jumps, respectively.

\begin{proposition}
\label{prop-1}
Consider the system $\mathcal{H}^{\Delta}_{\mathcal{M}}$ with $\Delta<\infty$, and $\mathcal{W}\subset\mathbb{R}^{n}$ is a closed set. If there exists a locally Lipschitz function $V: \mathbb{R}^{n}\rightarrow\mathbb{R}_{\geq0}$, $\alpha_{1}, \alpha_{2}\in\mathcal{K}_{\infty}$, $\rho\in\mathcal{PD}$, continuous nondecreasing functions $\gamma_{1}, \gamma_{2}: \mathbb{R}_{\geq0}\rightarrow\mathbb{R}_{\geq0}$ such that (A.1) and (B.2) hold, $\gamma_{1}$ satisfies the small gain condition and
\begin{enumerate}[(C.1)]
  \item  for all $(\varphi, u)\in\mathcal{C}$ and all $f\in\mathcal{F}(\varphi, u)$, $V(\varphi(0, 0))\geq\max\{\gamma_{1}(\bar{V}(\varphi)),\gamma_{2}(|u|)\}$ implies that $V^{\circ}(\varphi(0, 0), f)\leq0$;
  \item for arbitrary $\delta>0$, there exist $\gamma_{\delta}\in\mathcal{K}_{\infty}$ and $N_{\delta}>0$ such that for each solution $x$ with $\|\mathcal{A}^{\Delta}_{[0, 0]}x\|_{\mathcal{W}}\leq\delta$ and all $(t, j)\in\dom_{\geq0}x$, $t+j\geq T_{1}>0$ implies $t>\gamma_{\delta}(T_{1})-N_{\delta}$,
\end{enumerate}
then the set $\mathcal{W}$ is ISS for $\mathcal{H}^{\Delta}_{\mathcal{M}}$.
\end{proposition}

\begin{IEEEproof}
Following the similar fashion as the proof of Theorem \ref{thm-1}, for the first case that $V(\varphi(0, 0))\geq\max\{\gamma_{1}(\bar{V}(\varphi)),$ $\gamma_{2}(|u|)\}$, the conditions (B.2) and (C.1) are rewritten as
\begin{enumerate}[(i)]
  \item  $V^{\circ}(\varphi(0, 0), f)\leq0$ for all $(\varphi, u)\in\mathcal{C}$ and $f\in\mathcal{F}(\varphi, u)$;
  \item $V(g)-V(\varphi(0, 0))\leq-\rho(|\varphi(0, 0)|_{\mathcal{W}})$ for all $(\varphi, u)\in\mathcal{D}$ and $g\in\mathcal{G}(\varphi, u)$.
\end{enumerate}
Combining Proposition 3.24 in \cite{Goebel2012hybrid}, Theorem 3.40 in \cite{Goebel2012hybrid}, and the conditions (i)-(ii) and (C.2), there exists $\beta\in\mathcal{KLL}$ such that
\begin{equation*}
V(x(t, j))\leq\beta(\bar{V}_{[0, 0]}(x), t, j),  \quad \forall (t, j)\in\dom_{\geq0}x.
\end{equation*}
The remaining is the same as the proof of Theorem \ref{thm-1} and the set $\mathcal{W}$ is ISS for $\mathcal{H}^{\Delta}_{\mathcal{M}}$. Therefore, the proof is completed.
\end{IEEEproof}

\begin{proposition}
\label{prop-2}
Consider the system $\mathcal{H}^{\Delta}_{\mathcal{M}}$. Let $\Delta<\infty$ and $\mathcal{W}\subset\mathbb{R}^{n}$ be closed. If there exist a locally Lipschitz function $V: \mathbb{R}^{n}\rightarrow\mathbb{R}_{\geq0}$, $\alpha_{1}, \alpha_{2}\in\mathcal{K}_{\infty}$, $\rho\in\mathcal{PD}$, continuous nondecreasing functions $\gamma_{1}, \gamma_{2}: \mathbb{R}_{\geq0}\rightarrow\mathbb{R}_{\geq0}$ such that (A.1) and (B.1) hold, $\gamma_{1}$ satisfies the small gain condition and
\begin{enumerate}[(D.1)]
  \item  for all $(\varphi, u)\in\mathcal{D}$ and all $g\in\mathcal{G}(\varphi, u)$, $V(\varphi(0, 0))\geq\max\{\gamma_{1}(\bar{V}(\varphi)), \gamma_{2}(|u|)\}$ implies that $V(g)\leq V(\varphi(0, 0))$;
  \item for arbitrary $\delta>0$, there exist $\gamma_{\delta}\in\mathcal{K}_{\infty}$ and $N_{\delta}>0$ such that for each solution $x$ with $\|\mathcal{A}^{\Delta}_{[0, 0]}x\|_{\mathcal{W}}\leq\delta$ and $(t, j)\in\dom_{\geq0}x$, $t+j\geq T_{2}>0$ implies $j>\gamma_{\delta}(T_{2})-N_{\delta}$,
\end{enumerate}
then the set $\mathcal{W}$ is ISS for $\mathcal{H}^{\Delta}_{\mathcal{M}}$.
\end{proposition}

\begin{IEEEproof}
For the first case, the conditions (B.1) and (D.1) are rewritten as
\begin{enumerate}[(a)]
  \item for all $(\varphi, u)\in\mathcal{C}$ and $f\in\mathcal{F}(\varphi, u)$, $V^{\circ}(\varphi(0, 0), f)\leq-\rho(|\varphi(0, 0)|_{\mathcal{W}})$;
  \item  for all $(\varphi, u)\in\mathcal{D}$ and $g\in\mathcal{G}(\varphi, u)$, $V(g)\leq V(\varphi(0, 0))$.
\end{enumerate}
According to Theorem 2 in \cite{Liu2016razumikhin}, we obtain from (a)-(b) and (D.2) that there exists $\beta\in\mathcal{KLL}$ such that
\begin{equation*}
V(x(t, j))\leq\beta(\bar{V}_{[0, 0]}(x), t, j),  \quad \forall (t, j)\in\dom_{\geq0}x.
\end{equation*}
The remaining follows the same line as the proof of Theorem \ref{thm-1} and the set $\mathcal{W}$ is ISS for $\mathcal{H}^{\Delta}_{\mathcal{M}}$.
\end{IEEEproof}

\begin{remark}
\label{rmk-3}
Let us examine the differences between Theorems \ref{thm-1}-\ref{thm-2} and Propositions \ref{prop-1}-\ref{prop-2}. In Theorems \ref{thm-1}-\ref{thm-2}, LRFs decrease strictly during flow and at jumps. However, the flow in Proposition \ref{prop-2} and the jumps in Proposition \ref{prop-3} are persistent, which implies that the corresponding flow and jumps are neutral in stability analysis. Therefore, (C.2) and (D.2) are needed to ensure the convergence of the system state.
\hfill $\square$
\end{remark}

In the following, based on average dwell-time like condition, we study the unstable jump case and the unstable flow case, and obtain Razumikhin-type stability conditions.

\begin{theorem}
\label{thm-3}
Consider the system $\mathcal{H}^{\Delta}_{\mathcal{M}}$ with $\Delta<\infty$, and $\mathcal{W}\subset\mathbb{R}^{n}$ is a closed set. If there exist a locally Lipschitz function $V: \mathbb{R}^{n}\rightarrow\mathbb{R}_{\geq0}$, $\alpha_{1}, \alpha_{2}\in\mathcal{K}_{\infty}$, continuous nondecreasing functions $\gamma_{1}, \gamma_{2}: \mathbb{R}_{\geq0}\rightarrow\mathbb{R}_{\geq0}$ and constants $\lambda_{1}>0, \mu\geq1$, such that (A.1) holds, $\gamma_{1}$ satisfies the small gain condition and
\begin{enumerate}[(E.1)]
  \item for all $(\varphi, u)\in\mathcal{C}$ and all $f\in\mathcal{F}(\varphi, u)$, $V(\varphi(0, 0))\geq\max\{\gamma_{1}(\bar{V}(\varphi)), \gamma_{2}(|u|)\}$ implies that $V^{\circ}(\varphi(0, 0), f)\leq-\lambda_{1}V(\varphi(0, 0))$;
  \item for all $(\varphi, u)\in\mathcal{D}$ and all $g\in\mathcal{G}(\varphi, u)$, $V(\varphi(0, 0))\geq\max\{\gamma_{1}(\bar{V}(\varphi)), \gamma_{2}(|u|)\}$ implies that $V(g(\varphi, u))\leq\mu V(\varphi(0, 0))$;
  \item there exist $\varepsilon>0$ and $N_{0}\in\mathbb{Z}_{>0}$ such that the following holds:
  \begin{itemize}
    \item the average dwell-time like condition holds, that is, $\lambda_{1}\varepsilon-\ln\mu>0$,
    \item $j\in[\varepsilon^{-1}t-N_{0}, \varepsilon^{-1}t+N_{0}]$ holds for all $(t, j)\in\dom_{\geq0} x$,
  \end{itemize}
\end{enumerate}
then the set $\mathcal{W}$ is ISS for $\mathcal{H}^{\Delta}_{\mathcal{M}}$.
\end{theorem}

\begin{IEEEproof}
For the first case, for any $(t, j)\in[(t_{j}, j), (t_{j+1}, j+1))$, we obtain from (E.1) that
\begin{align}
\label{eqn-14}
V(x(t, j))\leq e^{-\lambda_{1}(t-t_{j})}V(x(t_{j}, j)).
\end{align}
At the jump time instant $(t_{j+1}, j+1)$, we have from (E.2) that
\begin{align}
\label{eqn-15}
V(x(t_{j+1}, j+1))&\leq\mu V(x(t_{j+1}, j)) \nonumber \\
&\leq\mu e^{-\lambda_{1}(t_{j+1}-t_{j})}V(x(t_{j}, j)).
\end{align}
According to \eqref{eqn-14}-\eqref{eqn-15} and using the mathematical induction, we conclude that for all $(t, j)\in\dom_{\geq0}x$
\begin{align}
\label{eqn-16}
V(x(t, j))&\leq\mu^{j}e^{-\lambda_{1}t}\alpha_{2}(\|\mathcal{A}^{\Delta}_{[0, 0]}x\|_{\mathcal{W}}).
\end{align}
Combining \eqref{eqn-16} and (E.3), we obtain that
\begin{align*}
V(x(t, j))&\leq e^{N_{0}\ln\mu} e^{0.5(\varepsilon^{-1}\ln\mu+\lambda_{1})(t+\varepsilon j)}\alpha_{2}(\|\mathcal{A}^{\Delta}_{[0, 0]}x\|_{\mathcal{W}})\\
&=:\beta(\|\mathcal{A}^{\Delta}_{[0, 0]}x\|_{\mathcal{W}}, t, j).
\end{align*}
Obviously, $\beta\in\mathcal{KLL}$.

The remaining is the same as the proof of Theorem \ref{thm-1} and the set $\mathcal{W}$ is ISS for $\mathcal{H}^{\Delta}_{\mathcal{M}}$. Therefore, the proof is completed.
\end{IEEEproof}

\begin{theorem}
\label{thm-4}
Consider the system $\mathcal{H}^{\Delta}_{\mathcal{M}}$ with $\Delta<\infty$ and $\mathcal{W}\subset\mathbb{R}^{n}$ is a closed set. If there exist a locally Lipschitz function $V: \mathbb{R}^{n}\rightarrow\mathbb{R}_{\geq0}$, $\alpha_{1}, \alpha_{2}\in\mathcal{K}_{\infty}$, continuous nondecreasing functions $\gamma_{1}, \gamma_{2}: \mathbb{R}_{\geq0}\rightarrow\mathbb{R}_{\geq0}$ and constants $\lambda_{1}>0, \mu\in(0, 1)$, such that (A.1) holds, $\gamma_{1}$ satisfies the small gain condition and
\begin{enumerate}[(F.1)]
  \item for all $(\varphi, u)\in\mathcal{C}$ and all $f\in\mathcal{F}(\varphi, u)$, if $V(\varphi(0, 0))\geq\max\{\gamma_{1}(\bar{V}(\varphi)), \gamma_{2}(|u|)\}$, then $V^{\circ}(\varphi(0, 0), f)\leq\lambda_{1}V(\varphi(0, 0))$;
  \item for all $(\varphi, u)\in\mathcal{D}$ and all $g\in\mathcal{G}(\varphi, u)$, $V(\varphi(0, 0))\geq\max\{\gamma_{1}(\bar{V}(\varphi)), \gamma_{2}(|u|)\}$ implies that $V(g)\leq\mu V(\varphi(0, 0))$;
  \item there exist $\varepsilon>0$ and $N_{0}\in\mathbb{Z}_{>0}$ such that the following holds:
  \begin{itemize}
    \item the reverse average dwell-time like condition holds, that is, $\lambda_{1}\varepsilon+\ln\mu<0$,
    \item $j\in[\varepsilon^{-1}t-N_{0}, \varepsilon^{-1}t+N_{0}]$ holds for all $(t, j)\in\dom_{\geq0} x$,
  \end{itemize}
\end{enumerate}
then the set $\mathcal{W}$ is ISS for $\mathcal{H}^{\Delta}_{\mathcal{M}}$.
\end{theorem}

The proof of Theorem \ref{thm-4} is similar to that of Theorem \ref{thm-3}, and hence omitted here. Theorems \ref{thm-3} and \ref{thm-4} generalize the results for typical hybrid systems with memory like impulsive time-delay systems \cite{Liu2011input, Ren2017stability}. In Theorems \ref{thm-3} and \ref{thm-4}, $\varepsilon>0$ and $N_{0}\in\mathbb{Z}_{>0}$ respectively play similar roles as (reverse) average dwell-time and chatter bound; see \cite{Liu2016lyapunov, Ren2016stability}. In addition, the condition that $j\in[\varepsilon^{-1}t-N_{0}, \varepsilon^{-1}t+N_{0}]$ is not strict. For instance, $\varepsilon$ exists in networked control systems due to hardware constraints and $t\geq \varepsilon j$ holds; see \cite{Heemels2010networked} for more details.

\section{ISS with Lyapunov-Krasovskii Functionals}
\label{sec-krasovskii}

Besides Lyapunov-Razumikhin functions, Lyapunov-Krasovskii functionals are another extension of classic Lyapunov functions. For Lyapunov-Krasovskii functional $V: \mathcal{M}^{\Delta}\rightarrow\mathbb{R}_{\geq0}$, the upper Dini derivative $D^{+}V$ of $V$ at $\varphi\in\mathcal{M}^{\Delta}$ along the solutions of $\mathcal{H}^{\Delta}_{\mathcal{M}}$ is given by
\begin{equation*}
D^{+}V(\varphi):=\sup_{(x, u)\in\mathfrak{S}^{\Delta}_{u}(\varphi), \mathcal{A}^{\Delta}_{[0, 0]}x=\varphi}\limsup_{h\rightarrow0^{+}}\frac{V(\mathcal{A}^{\Delta}_{[h, 0]}x)-V(\varphi)}{h}.
\end{equation*}
In this section, the upper Dini derivative of LKFs is implemented. Next, we first establish Krasovskii-type conditions for ISS of the hybrid system \eqref{eqn-1}, then provide two relaxations, and finally extend the obtained results to the cases that the LKFs are not strictly decreasing during the flow or at the jumps.

\subsection{Krasovskii-type Stability Conditions}

\begin{theorem}
\label{thm-5}
Consider the system $\mathcal{H}^{\Delta}_{\mathcal{M}}$. Let $\Delta<\infty$ and $\mathcal{W}\subset\mathbb{R}^{n}$ be closed. If there exist a functional $V: \mathcal{M}^{\Delta}\rightarrow\mathbb{R}_{\geq0}$, $\alpha_{1}, \alpha_{2}, \rho\in\mathcal{K}_{\infty}$, $\alpha_{3}\in\mathcal{PD}$ such that
\begin{enumerate}[(G.1)]
  \item  for all $\varphi\in\mathbb{R}^{n}$, $\alpha_{1}(|\varphi(0, 0)|_{\mathcal{W}})\leq V(\varphi)\leq\alpha_{2}(|\varphi|_{\mathcal{W}})$;
  \item  for all $(\varphi, u)\in\mathcal{C}$, $V(\varphi)\geq\rho(|u|)$ implies that $D^{+}V(\varphi)\leq-\alpha_{3}(|\varphi(0, 0)|_{\mathcal{W}})$;
  \item  for all $(\varphi, u)\in\mathcal{D}$ and all $g\in\mathcal{G}(\varphi, u)$, $V(\varphi)\geq\rho(|u|)$ implies that $V(\varphi^{+}_{g})\leq V(\varphi)-\alpha_{3}(|\varphi(0, 0)|_{\mathcal{W}})$,
\end{enumerate}
where $\varphi^{+}_{g}$ is the hybrid memory arc following $\varphi$ after a jump of value $g\in\mathcal{G}$, then the set $\mathcal{W}$ is ISS for $\mathcal{H}^{\Delta}_{\mathcal{M}}$.
\end{theorem}

\begin{IEEEproof}
According to (G.2)-(G.3), the proof is divided into two cases: the first case that $V(\varphi)\geq\rho(|u|)$ and the second case that $V(\varphi)<\rho(|u|)\}$.

For the first case, the conditions (G.2)-(G.3) are written as
\begin{align}
\label{eqn-17}
D^{+}V(\varphi)&\leq-\alpha_{3}(|\varphi(0, 0)|_{\mathcal{W}}), \\
\label{eqn-18}
V(\varphi^{+}_{g})&\leq V(\varphi)-\alpha_{3}(|\varphi(0, 0)|_{\mathcal{W}}).
\end{align}
Based on the proof of Theorem 2 in \cite{Liu2016lyapunov}, it follows from \eqref{eqn-17}-\eqref{eqn-18} that there exists $\beta\in\mathcal{KLL}$ such that
\begin{equation}
\label{eqn-19}
V(\mathcal{A}^{\Delta}_{[t, j]}x)\leq\beta(\|\mathcal{A}^{\Delta}_{[0, 0]}x\|_{\mathcal{W}}, t, j),  \quad \forall (t, j)\in\dom_{\geq0}x.
\end{equation}
It obtains from \eqref{eqn-19} and the second case that for all $(t, j)\in\dom_{\geq0}x$,
\begin{equation}
\label{eqn-20}
V(\mathcal{A}^{\Delta}_{[t, j]}x)\leq\max\{\beta(\|\mathcal{A}^{\Delta}_{[0, 0]}x\|_{\mathcal{W}}, t, j), \rho(\|u\|_{(t, j)})\}.
\end{equation}

Therefore, we get from (G.1) that for all $(t, j)\in\dom_{\geq0}x$,
\begin{align*}
|x(t, j)|_{\mathcal{W}}\leq\max\{\beta(\|\mathcal{A}^{\Delta}_{[0, 0]}x\|_{\mathcal{W}}, t, j), \gamma(\|u\|_{(t, j)})\},
\end{align*}
where $\beta(v, t, j):=2\alpha_{2}(\bar{\beta}(v, t, j))$ and $\gamma(v):=2\alpha_{2}(\rho(v))$. Hence, the set $\mathcal{W}$ is ISS for $\mathcal{H}^{\Delta}_{\mathcal{M}}$.
\end{IEEEproof}

\subsection{Extensions of Krasovskii-type Stability Conditions}

In the following, we extend Theorem \ref{thm-5} to some general cases, such as the case that $\mathcal{H}^{\Delta}_{\mathcal{M}}$ is strictly decreasing during flow but nonincreasing during jumps. Similar to Propositions \ref{prop-1}-\ref{prop-2}, the following relaxations of Theorem \ref{thm-5} are established. Their proofs are combinations of the proof strategies of Propositions \ref{prop-1}-\ref{prop-2} and Theorem \ref{thm-5}, and hence omitted here.

\begin{proposition}
\label{prop-3}
Consider the system $\mathcal{H}^{\Delta}_{\mathcal{M}}$. Let $\Delta<\infty$ and $\mathcal{W}\subset\mathbb{R}^{n}$ be closed. If there exist a functional $V: \mathcal{M}^{\Delta}\rightarrow\mathbb{R}_{\geq0}$, $\alpha_{1}, \alpha_{2}, \rho\in\mathcal{K}_{\infty}$ and $\alpha_{3}\in\mathcal{PD}$ such that (C.2) and (G.1) hold, and
\begin{align*}
&V(\varphi)\geq\rho(|u|) \\
&  \Rightarrow\left\{\begin{aligned}
&D^{+}V(\varphi)\leq0, \ (\varphi, u)\in\mathcal{C},   \\
&V(\varphi^{+}_{g})\leq V(\varphi)-\alpha_{3}(|\varphi(0, 0)|_{\mathcal{W}}), \ (\varphi, u)\in\mathcal{D}, g\in\mathcal{G}(\varphi, u),
\end{aligned}\right.
\end{align*}
then the set $\mathcal{W}$ is ISS for $\mathcal{H}^{\Delta}_{\mathcal{M}}$.
\end{proposition}

\begin{proposition}
\label{prop-4}
Consider the system $\mathcal{H}^{\Delta}_{\mathcal{M}}$. Let $\Delta<\infty$ and $\mathcal{W}\subset\mathbb{R}^{n}$ be closed. If there exist a functional $V: \mathcal{M}^{\Delta}\rightarrow\mathbb{R}_{\geq0}$, $\alpha_{1}, \alpha_{2}, \rho\in\mathcal{K}_{\infty}$ and $\alpha_{3}\in\mathcal{PD}$ such that (D.2) and (G.1) hold, and
\begin{align*}
&V(\varphi)\geq\rho(|u|) \\
& \Rightarrow \left\{\begin{aligned}
&D^{+}V(\varphi)\leq-\alpha_{3}(|\varphi(0, 0)|_{\mathcal{W}}), \quad (\varphi, u)\in\mathcal{C},   \\
&V(\varphi^{+}_{g})\leq V(\varphi), \quad (\varphi, u)\in\mathcal{D}, g\in\mathcal{G}(\varphi, u),
\end{aligned}\right.
\end{align*}
then the set $\mathcal{W}$ is ISS for $\mathcal{H}^{\Delta}_{\mathcal{M}}$.
\end{proposition}

The following results, which are based on average dwell-time like conditions, establish Krasovskii-type stability conditions for $\mathcal{H}^{\Delta}_{\mathcal{M}}$ in both the stable flow case and the stable jump case, respectively.

\begin{theorem}
\label{thm-6}
Consider the system $\mathcal{H}^{\Delta}_{\mathcal{M}}$. Let $\Delta<\infty$ and $\mathcal{W}\subset\mathbb{R}^{n}$ be closed. If there exist a functional $V: \mathcal{M}^{\Delta}\rightarrow\mathbb{R}_{\geq0}$, $\alpha_{1}, \alpha_{2}, \rho\in\mathcal{K}_{\infty}$ and constants $\lambda_{1}>\lambda_{2}\geq0$, $\mu>1$ such that (G.1) holds, and
\begin{enumerate}[(H.1)]
  \item for all $(\varphi, u)\in\mathcal{C}$, $V(\varphi)\geq\rho(|u|)$ implies that $D^{+}V(\varphi)\leq-\lambda_{1}V(\varphi)+\lambda_{2}\hat{V}(\varphi)$;
  \item for all $(\varphi, u)\in\mathcal{D}$ and all $g\in\mathcal{G}(\varphi, u)$, $V(\varphi)\geq\rho(|u|)$ implies that $V(\varphi^{+}_{g})\leq\mu V(\varphi)$;
  \item there exist $\varepsilon>0$ and $N_{0}\in\mathbb{Z}_{>0}$ such that the following holds:
  \begin{itemize}
    \item $\varepsilon\lambda-\ln\mu>0$, where $\lambda\in(0, \bar{\lambda})$ and $\bar{\lambda}$ is the solution to $\bar{\lambda}-\lambda_{1}+\lambda_{2}e^{\bar{\lambda}(\Delta+1)}=0$,
    \item $j\in[\varepsilon^{-1}t-N_{0}, \varepsilon^{-1}t+N_{0}]$ holds for all $(t, j)\in\dom_{\geq0} x$,
  \end{itemize}
\end{enumerate}
where $\hat{V}(\mathcal{A}^{\Delta}_{[t, j]}x):=\sup_{s+k\in[-\Delta-1, 0]}V(\mathcal{A}^{\Delta}_{[t+s, j+k]}x)$ if $(t+s, j+k)\in\dom_{\geq0}x$; otherwise, $\hat{V}=V$, then the set $\mathcal{W}$ is ISS for $\mathcal{H}^{\Delta}_{\mathcal{M}}$.
\end{theorem}

\begin{IEEEproof}
The proof is divided into the following three parts. First, the existence of $\bar{\lambda}$ in (H.3) is determined. Second, the boundedness of $V$ is established. Finally, according to (H.3) and the bound of $V$, the convergence of the system state is guaranteed.

\textbf{Part 1.} Define $\Gamma(\lambda):=\lambda-\lambda_{1}+\lambda_{2}e^{\lambda(\Delta+1)}$. Since $\lambda_{1}>\lambda_{2}\geq0$, $\Gamma(0)<0$. Moreover, $\Gamma(\lambda)\rightarrow\infty$ as $\lambda\rightarrow\infty$. In addition, $\Gamma'(\lambda):=1+(\Delta+1)\lambda_{2}e^{\lambda(\Delta+1)}>0$. Therefore, there exists a unique $\bar{\lambda}>0$ such that $\Gamma(\bar{\lambda})=0$, and $\Gamma(\lambda)<0$ for all $\lambda\in[0, \bar{\lambda})$. 

\textbf{Part 2.} Define $W(\mathcal{A}^{\Delta}_{[t, j]}x):=e^{\lambda t}V(\mathcal{A}^{\Delta}_{[t, j]}x)$, where $\lambda$ is defined in (H.3). Note that for each $(t, j)\in\dom_{\geq0}x$, $(t, j)$ satisfies $(t_{j}, j)\preceq(t, j)\prec(t_{j+1}, j+1)$. In the following, we prove via contradiction that for all $(t, j)\in\dom x$ and $V(\varphi)\geq\rho(|u|)$,
\begin{align}
\label{eqn-21}
W(\mathcal{A}^{\Delta}_{[t, j]}x)&\leq\mu^{j}\alpha_{2}(\|\mathcal{A}^{\Delta}_{[0, 0]}x\|_{\mathcal{W}})=:\mathfrak{U}(t, j).
\end{align}

If \eqref{eqn-21} does not hold, then there are two cases as follows. The first case is that $W(\mathcal{A}^{\Delta}_{[t, j]}x)$ jumps such that \eqref{eqn-21} is invalid. That is, there exist $(t_{k+1}, k)\in\dom_{\geq0}x$ and $k\in\mathbb{Z}_{\geq0}$ such that
\begin{align}
\label{eqn-22}
W(\mathcal{A}^{\Delta}_{[t, j]}x)&\leq\mathfrak{U}(t, j), \quad (t, j)\prec(t_{k+1}, k+1); \\
\label{eqn-23}
W(\mathcal{A}^{\Delta}_{[t_{k+1}, k+1]}x)&>\mathfrak{U}(t_{k+1}, k+1).
\end{align}
However, it follows from (H.2) and \eqref{eqn-21} that
\begin{align*}
W(\mathcal{A}^{\Delta}_{[t_{k+1}, k+1]}x)&\leq\mu W(\mathcal{A}^{\Delta}_{[t_{k+1}, k]}x)\\
&\leq\mu^{k+1}\alpha_{2}(\|\mathcal{A}^{\Delta}_{[0, 0]}x\|_{\mathcal{W}})=\mathfrak{U}(t_{k+1}, k+1),
\end{align*}
which contradicts with \eqref{eqn-23}. Thus, \eqref{eqn-21} holds for all the jumps.

The second case is $W(\mathcal{A}^{\Delta}_{[t, j]}x)$ flows such that \eqref{eqn-21} fails. In the sequel, there exists $(T, N)\in\dom_{\geq0}x$ with $(t_{k}, k)\preceq(T, N)\prec(t_{k+1}, k)$, $k\in\mathbb{Z}_{\geq0}$, such that
\begin{align}
\label{eqn-24}
W(\mathcal{A}^{\Delta}_{[t, j]}x)&\leq\mathfrak{U}(t, j), \quad \forall (t, j)\preceq(T, N); \\
\label{eqn-25}
W(\mathcal{A}^{\Delta}_{[T, N]}x)&=\mathfrak{U}(T, N); \\
\label{eqn-26}
W(\mathcal{A}^{\Delta}_{[t, N]}x)&>\mathfrak{U}(t, N), \quad t\in(T, T+\Delta t),
\end{align}
where $\Delta t>0$ is arbitrarily small. It obtains from \eqref{eqn-24}-\eqref{eqn-26} that
\begin{align}
\label{eqn-27}
D^{+}W(\mathcal{A}^{\Delta}_{[T, N]}x)\geq D^{+}\mathfrak{U}(T, N)=0.
\end{align}
On the other hand, it follows from (H.1) and (H.3) that
\begin{align}
\label{eqn-28}
D^{+}W(\mathcal{A}^{\Delta}_{[T, N]}x)&=\lambda W(\mathcal{A}^{\Delta}_{[T, N]}x)+e^{\lambda T}D^{+}V(\mathcal{A}^{\Delta}_{[T, N]}x) \nonumber \\
&\leq\lambda W(\mathcal{A}^{\Delta}_{[T, N]}x)+e^{\lambda T}[-\lambda_{1}V(\mathcal{A}^{\Delta}_{[T, N]}x) \nonumber \\
&\quad  +\lambda_{2}\hat{V}(\mathcal{A}^{\Delta}_{[T, N]}x)].
\end{align}
If $(T+s, N+k)\notin\dom_{\geq0}x$ for $s+k\in[-\Delta-1, 0]$, it follows from the definition of $\hat{V}$ that $e^{\lambda T}\hat{V}(\mathcal{A}^{\Delta}_{[T, N]}x))=e^{\lambda T}V(\mathcal{A}^{\Delta}_{[T, N]}x))=W(\mathcal{A}^{\Delta}_{[T, N]}x)$, which implies that $\lambda_{2}e^{\lambda T}\hat{V}(\mathcal{A}^{\Delta}_{[T, N]}x))\leq\lambda_{2}e^{\lambda(\Delta+1)}W(\mathcal{A}^{\Delta}_{[T, N]}x)$. If $(T+s, N+k)\in\dom_{\geq0}x$, then we get from the definition of $\hat{V}$ and \eqref{eqn-24}-\eqref{eqn-25} that
\begin{align}
\label{eqn-29}
&\lambda_{2}e^{\lambda T}\hat{V}(\mathcal{A}^{\Delta}_{[T, N]}x)) \nonumber \\
&=\lambda_{2}e^{\lambda T}\sup_{s+k\in[-\Delta-1, 0]}V(\mathcal{A}^{\Delta}_{[T+s, N+k]}x)  \nonumber\\
&\leq\lambda_{2}\sup_{s+k\in[-\Delta-1, 0]}e^{\lambda(T+\Delta+1+s)}V(\mathcal{A}^{\Delta}_{[T+s, N+k]}x) \nonumber\\
&=\lambda_{2}e^{\lambda(\Delta+1)}\sup_{s+k\in[-\Delta-1, 0]}e^{\lambda(T+s)}V(\mathcal{A}^{\Delta}_{[T+s, N+k]}x) \nonumber\\
&=\lambda_{2}e^{\lambda(\Delta+1)}\sup_{s+k\in[-\Delta-1, 0]}W(\mathcal{A}^{\Delta}_{[T+s, N+k]}x) \nonumber\\
&\leq\lambda_{2}e^{\lambda(\Delta+1)}\sup_{s+k\in[-\Delta-1, 0]}\mathfrak{U}(T+s, N+k) \nonumber\\
&\leq\lambda_{2}e^{\lambda(\Delta+1)}\mathfrak{U}(T, N) \nonumber\\
&=\lambda_{2}e^{\lambda(\Delta+1)}W(\mathcal{A}^{\Delta}_{[T, N]}x).
\end{align}
Therefore, we have from \eqref{eqn-28} and \eqref{eqn-29} that
\begin{align}
\label{eqn-30}
D^{+}W(\mathcal{A}^{\Delta}_{[T, N]}x)&\leq\lambda W(\mathcal{A}^{\Delta}_{[T, N]}x)-\lambda_{1}W(\mathcal{A}^{\Delta}_{[T, N]}x) \nonumber  \\
&\quad +\lambda_{2}e^{\lambda(\Delta+1)}W(\mathcal{A}^{\Delta}_{[T, N]}x)<0,
\end{align}
which contradicts with \eqref{eqn-27}. Thus, \eqref{eqn-21} holds in the flow.

\textbf{Part 3.} Based on above analysis and using the mathematical induction, we obtain that if $V(\varphi)\geq\rho(|u|)$, then \eqref{eqn-21} holds for all $(t, j)\in\dom_{\geq0}x$. Combining \eqref{eqn-21} and the case that $V(\varphi)\leq\rho(|u|)$ yields that for all $(t, j)\in\dom_{\geq0}x$,
\begin{align*}
&V(\mathcal{A}^{\Delta}_{[t, j]}x)\leq\max\{e^{-\lambda t}\mu^{j}\alpha_{2}(\|\mathcal{A}^{\Delta}_{[0, 0]}x\|_{\mathcal{W}}), \rho(\|u\|_{(t, i)})\}  \nonumber \\
&\qquad \leq\max\{e^{\pi} e^{0.5\theta(t+\varepsilon j)}\alpha_{2}(\|\mathcal{A}^{\Delta}_{[0, 0]}x\|_{\mathcal{W}}), \rho(\|u\|_{(t, j)})\},
\end{align*}
where $\pi:=0.5(\ln\mu+\lambda\varepsilon)N_{0}$ and $\theta:=\varepsilon^{-1}\ln\mu-\lambda<0$ from (H.3). The remaining is similar to the proof of Theorem \ref{thm-5}, and the set $\mathcal{W}$ is ISS for $\mathcal{H}^{\Delta}_{\mathcal{M}}$.
\end{IEEEproof}

\begin{theorem}
\label{thm-7}
Consider the system $\mathcal{H}^{\Delta}_{\mathcal{M}}$. Let $\Delta<\infty$ and $\mathcal{W}\subset\mathbb{R}^{n}$ be closed. If there exist a functional $V: \mathcal{M}^{\Delta}\rightarrow\mathbb{R}_{\geq0}$, $\alpha_{1}, \alpha_{2}, \rho\in\mathcal{K}_{\infty}$ and constants $\lambda_{1}, \lambda_{2}\geq0$, $\mu\in(0, 1)$ such that (G.1) holds, and
\begin{enumerate}[({I}.1)]
  \item for all $(\varphi, u)\in\mathcal{C}$, $V(\varphi)\geq\rho(|u|)$ implies that $D^{+}V(\varphi)\leq\lambda_{1}V(\varphi)+\lambda_{2}\hat{V}(\varphi)$;
  \item for all $(\varphi, u)\in\mathcal{D}$ and all $g\in\mathcal{G}(\varphi, u)$, $V(\varphi)\geq\rho(|u|)$ implies that $V(\varphi^{+}_{g})\leq\mu V(\varphi)$;
  \item there exist $\varepsilon>0$ and $N_{0}\in\mathbb{Z}_{>0}$ such that the following holds:
  \begin{itemize}
    \item $\ln\mu+\varepsilon\lambda<0$, where $\lambda=\lambda_{1}+\lambda_{2}\mu^{-N_{0}}e^{(\Delta+1)}$,
    \item $j\in[\varepsilon^{-1}t-N_{0}, \varepsilon^{-1}t+N_{0}]$ holds for all $(t, j)\in\dom_{\geq0} x$,
  \end{itemize}
\end{enumerate}
then the set $\mathcal{W}$ is ISS for $\mathcal{H}^{\Delta}_{\mathcal{M}}$.
\end{theorem}

\begin{IEEEproof}
Similar to the proof of Theorem \ref{thm-6}, the proof is partitioned into the following two steps.

\textbf{Step 1.} We consider the first case that $V(\varphi)\geq\rho(|u|)$. In this case, we prove that for all $(t, j)\in\dom_{\geq0}x$, 
\begin{align}
\label{eqn-31}
V(\mathcal{A}^{\Delta}_{[t, j]}x)\leq\mu^{j}e^{\lambda t}\alpha_{2}(\|\mathcal{A}^{\Delta}_{[0, 0]}x\|_{\mathcal{W}}).
\end{align}
Similar to the proof of Theorem \ref{thm-7}, if \eqref{eqn-31} does not hold, then there are following two scenarios. The first scenario is that $W(\mathcal{A}^{\Delta}_{[t, j]}x)$ flows such that \eqref{eqn-31} fails. That is, there exists $(t_{k}, k)\preceq(T, N)\prec(t_{k+1}, k)$ such that
\begin{align}
\label{eqn-32}
V(\mathcal{A}^{\Delta}_{[t, j]}x)&\leq\mu^{j}e^{\lambda t}\alpha_{2}(\|\mathcal{A}^{\Delta}_{[0, 0]}x\|_{\mathcal{W}}), \ \forall (t, j)\prec(T, N),  \\
\label{eqn-33}
V(\mathcal{A}^{\Delta}_{[T, N]}x)&=\mu^{N}e^{\lambda T}\alpha_{2}(\|\mathcal{A}^{\Delta}_{[0, 0]}x\|_{\mathcal{W}}),  \\
\label{eqn-34}
V(\mathcal{A}^{\Delta}_{[t, N]}x)&>\mu^{N}e^{\lambda t}\alpha_{2}(\|\mathcal{A}^{\Delta}_{[0, 0]}x\|_{\mathcal{W}}), \ t\in(T, T+\Delta t),
\end{align}
where $\Delta t>0$ is arbitrarily small. As a result,
\begin{equation}
\label{eqn-35}
D^{+}V(\mathcal{A}^{\Delta}_{[T, N]}x)\geq\lambda\mu^{N}e^{\lambda T}\alpha_{2}(\|\mathcal{A}^{\Delta}_{[0, 0]}x\|_{\mathcal{W}}).
\end{equation}
On the other hand, we have from \eqref{eqn-29} and (I.1) that
\begin{align*}
&\lambda\mu^{N}e^{\lambda T}\alpha_{2}(\|\mathcal{A}^{\Delta}_{[0, 0]}x\|_{\mathcal{W}}) \\
&=(\lambda_{1}+\lambda_{2}\mu^{-N_{0}}e^{(\Delta+1)})V(\mathcal{A}^{\Delta}_{[T, N]}x) \\
&\geq\lambda_{1}V(\mathcal{A}^{\Delta}_{[T, N]}x)+\lambda_{2}\mu^{-N_{0}}\hat{V}(\mathcal{A}^{\Delta}_{[T, N]}x)\\
&>D^{+}V(\mathcal{A}^{\Delta}_{[T, N]}x),
\end{align*}
which contradicts with \eqref{eqn-35}. Hence, \eqref{eqn-31} holds during the flow.

The second scenario is that $W(\mathcal{A}^{\Delta}_{[t, j]})$ jumps such that \eqref{eqn-31} fails. In this scenario, we have
\begin{align*}
V(\mathcal{A}^{\Delta}_{[t, j+1]}x)&\leq\mu V(\mathcal{A}^{\Delta}_{[t, j]}x)\leq\mu^{j+1}e^{\lambda t}\alpha_{2}(\|\mathcal{A}^{\Delta}_{[0, 0]}x\|_{\mathcal{W}}),
\end{align*}
which is a contradiction. Thus, \eqref{eqn-31} holds for all the jumps.

As a result, according to the mathematical induction, we have that \eqref{eqn-31} holds for all $(t, j)\in\dom_{\geq0}x$, which implies from (I.3) that
\begin{align*}
V(\mathcal{A}^{\Delta}_{[t, j]}x)&\leq e^{\pi}e^{0.5(\varepsilon^{-1}\ln\mu+\lambda)(t+\varepsilon j)}\alpha_{2}(\|\mathcal{A}^{\Delta}_{[0, 0]}x\|_{\mathcal{W}})\\
&=:\beta(\|\mathcal{A}^{\Delta}_{[0, 0]}x\|_{\mathcal{W}}, t, j),
\end{align*}
where $\pi:=-(1.5\ln\mu+0.5\lambda\varepsilon)N_{0}$.

\textbf{Step 2.} Combining the first case and the second case that $V(\varphi)\leq\rho(|u|)$ yields that for all $(t, j)\in\dom_{\geq0}x$,
\begin{align*}
V(\mathcal{A}^{\Delta}_{[t, j]}x)\leq\max\{\beta(\|\mathcal{A}^{\Delta}_{[0, 0]}x\|_{\mathcal{W}}, t, j), \rho(\|u\|_{(t, j)})\}.
\end{align*}
It follows from (G.1) that the set $\mathcal{W}$ is ISS for $\mathcal{H}^{\Delta}_{\mathcal{M}}$.
\end{IEEEproof}

\begin{remark}
\label{rmk-4}
Theorems \ref{thm-6} and \ref{thm-7} generalize the results for typical hybrid systems with memory, such as switched time-delay systems \cite{Liu2011input} and impulsive time-delay systems \cite{Chen2009input, Wu2016input, Dashkovskiy2012stability, Naghshtabrizi2010stability}. In addition, if $\lambda_{1}>\lambda_{2}\geq0$, $\mu\in(0, 1)$, then (H.3) is not needed. In this case, Theorem \ref{thm-6} is an extension of Corollary 1 in \cite{Liu2016lyapunov}.
\hfill $\square$
\end{remark}

\section{Illustrative Examples}
\label{sec-example}

In this section, two examples are presented to illustrate the developed results. The first example is motivated by networked control systems considered in \cite{Heemels2010networked}. The second example is modified from impulsive switched time-delay systems studied in \cite{Liu2011input}.

\textbf{Example 1.} Consider the system $\mathcal{H}^{\Delta}_{\mathcal{M}}=(\mathcal{F}, \mathcal{G}, \mathcal{C}, \mathcal{D}, \mathbb{R}^{n}, \mathbb{R}^{m})$ with the following data: $\varphi=(\varphi_{x}, \varphi_{e}, \varphi_{s}, \varphi_{\tau}, \varphi_{l})$,
\begin{align*}
\mathcal{C}&=\{\varphi\in\mathcal{M}^{\Delta}|(\varphi_{l}(0, 0)=0\wedge\varphi_{\tau}(0, 0)\in[0, \tau_{\mati}])\\
&\quad \vee(\varphi_{l}(0, 0)=1\wedge\varphi_{\tau}(0, 0)\in[0, \tau_{\mad}])\},  \\
\mathcal{D}&=\{\varphi\in\mathcal{M}^{\Delta}|(\varphi_{l}(0, 0)=0\wedge\varphi_{\tau}(0, 0)\in[\varepsilon, \tau_{\mati}])\\
&\quad \vee(\varphi_{l}(0, 0)=1\wedge\varphi_{\tau}(0, 0)\in[0, \tau_{\mad}])\}, \\
\mathcal{F}(\varphi, u)&=\begin{bmatrix}\begin{smallmatrix}
\bigcup\limits_{k\in J-r}-7\varphi_{x}(0, 0)+\varphi_{x}(-r, k)+\varphi_{e}(0, 0)+u(0, 0)  \\
\bigcup\limits_{k\in J-r}5\varphi_{x}(0, 0)-\varphi_{x}(-r, k)-\varphi_{e}(0, 0)+u(0, 0)  \\ 0  \\ 1 \\0
\end{smallmatrix}\end{bmatrix},
\end{align*}
and $\mathcal{G}(\varphi, u)=(\varphi_{x}(0, 0), \frac{1}{2}\varphi_{e}(0, 0), 0, 0, 1)^{\top}$ for $\varphi_{l}(0, 0)=0$; $\mathcal{G}(\varphi, u)=(\varphi_{x}(0, 0), \varphi_{e}(0, 0)+\varphi_{s}(0, 0), -\varphi_{e}(0, 0)-\varphi_{s}(0, 0), \varphi_{\tau}(0, 0), 0)^{\top}$ for $\varphi_{l}(0, 0)=1$. Moreover, $\Delta:=r+r/\varepsilon+1$, $r>0$ is a constant delay; $\epsilon$, $\tau_{\mad}$ and $\tau_{\mati}$ are constants satisfying $0<\epsilon<\tau_{\mati}\leq\bar{\tau}$ and $0\leq\tau_{\mad}\leq\tau_{\mati}$ with $\bar{\tau}=0.04125$.

To analyze stability of the set $\mathcal{W}:=\{0\}\times\{0\}\times\{0\}\times((\{0\}\times[\epsilon, \tau_{\mati}])\vee(\{1\}\times[0, \tau_{\mad}]))$, choose an LRF as $V(\varphi):=V_{1}(\varphi_{x})+\phi_{\varphi_{l}}(\varphi_{\tau})V^{2}_{2}(\varphi_{e}, \varphi_{s})$ where $V_{1}(\varphi_{x})=|\varphi_{x}|^{2}$, $V_{2}(\varphi_{e}, \varphi_{s})=|\varphi_{e}+\varphi_{s}|$ and $\phi_{\varphi_{l}}(\varphi_{\tau})$ are the solutions to the differential equations
\begin{align*}
\dot{\phi}_{0}&=-8\phi_{0}-\frac{64}{7}\phi^{2}_{0}-8, \quad \phi_{0}(0)=2, \\
\dot{\phi}_{1}&=-10\phi_{1}-\frac{64}{7}\phi^{2}_{1}-15, \quad \phi_{1}(0)=2.2.
\end{align*}
Since $\phi_{l}(\bar{\tau})\in[0.5, 2.2]$, $\phi_{l}(\tau)$ is strictly decreasing on $[0, \tau_{\mati}]$ and $\phi_{l}(\tau)\in[0.5, 2.2]$ for $\tau\in[0, \tau_{\mati}]$. In addition, assume that $\phi_{0}(\tau)\geq0.5\phi_{1}(0)$ for $\tau\in[0, \tau_{\mati}]$ and $\phi_{1}(\tau)\geq\phi_{0}(\tau)$ for $\tau\in[0, \tau_{\mad}]$. Hence, $\alpha_{1}(\varphi)=|\varphi_{x}|^{2}$ and $\alpha_{2}(\varphi)=|\varphi_{x}|^{2}+2.2|\varphi_{e}+\varphi_{s}|^{2}$.

Consider the LRF at the jumps. For all $\varphi$ with $\varphi_{\tau}(0, 0)\in[\epsilon, \tau_{\mati}]$, it holds that for $\varphi_{l}(0, 0)=0$,
\begin{align*}
V(\mathcal{G}(\varphi, u))&=V_{1}(\varphi_{x}(0, 0))+\phi_{0}(0)V^{2}_{2}(\varphi_{e}(0, 0)/2, 0)  \\
&\leq V_{1}(\varphi_{x}(0, 0))  \nonumber  \\
&\quad +\phi_{1}(\varphi_{\tau}(0, 0))V^{2}_{2}(\varphi_{e}(0, 0), \varphi_{s}(0, 0))  \\
&=V(\varphi(0, 0)),
\end{align*}
and for $\varphi_{l}(0, 0)=1$,
\begin{align*}
V(\mathcal{G}(\varphi, u))&=V_{1}(\varphi_{x}(0, 0))  \nonumber  \\
&\quad +\phi(\varphi_{\tau}(0, 0))V^{2}_{2}(\varphi_{e}(0, 0), \varphi_{s}(0, 0)) \\
&=V_{1}(\varphi_{x}(0, 0))\leq V(\varphi(0, 0)).
\end{align*}

Consider the LRF in the flow. For all $\varphi\in\mathcal{M}^{\Delta}$ with $\varphi_{\tau}(0, 0)\in[0, \tau_{\mati}]$, if $V(\varphi(0, 0))\geq\max\{ 0.5\bar{V}_{[0,0]}(\varphi), |u(0, 0)|^{2}\}$, then it follows from the fact $z_{1}z_{2}\leq az^{2}_{1}+z^{2}_{2}/a$ for any $a>0$ that, for $\varphi_{l}(0, 0)=0$,
\begin{align*}
&V^{\circ}(\varphi(0, 0), \mathcal{F}(\varphi, u))\leq-14|\varphi_{x}(0, 0)|^{2}+2|\varphi_{x}(0, 0)|\|\varphi_{x}\|_{\Delta}\\
&\quad +2|\varphi_{x}(0, 0)||\varphi_{e}(0, 0)|+2|\varphi_{x}(0, 0)||u(0, 0)|  \\
&\quad +\left(-8\phi_{0}(\varphi_{\tau}(0, 0))-\frac{64}{7}\phi^{2}_{0}(\varphi_{\tau}(0, 0))-8\right)|\varphi_{e}(0, 0)|^{2}  \\
&\quad +2\phi_{0}(\varphi_{\tau}(0, 0))|\varphi_{e}(0, 0)|(5|\varphi_{x}(0, 0)|+\|\varphi_{x}\|_{\Delta}\\
&\quad +|\varphi_{e}(0, 0)|+|u(0, 0)|) \\
&\leq-|\varphi_{x}(0, 0)|^{2}-|\varphi_{e}(0, 0)|^{2}.
\end{align*}
For $\varphi_{l}(0, 0)=1$, it follows from the same fashion that $V^{\circ}(\varphi(0, 0), f(\varphi, u))\leq-|\varphi_{x}(0, 0)|^{2}-|\varphi_{e}(0, 0)|^{2}$.

Based on the above analysis, all the conditions in Proposition \ref{prop-3} are satisfied. As a result, the set $\mathcal{W}=\{0\}\times\{0\}\times\{0\}\times(([\epsilon, \tau_{\mati}]\times\{0\})\vee([0, \tau_{\mad}]\times\{1\}))$ is ISS for $\mathcal{H}^{\Delta}_{\mathcal{M}}$.

\textbf{Example 2.} Consider linear impulsive switched time-delay systems of the form (see also \cite{Liu2011input})
\begin{align}
\label{eqn-36}
\left.\begin{aligned}
\dot{x}&=A_{p}x+B_{p}\hat{x}+C_{p}u \\
\dot{p}&=0 \\
\dot{\tau}&=1
\end{aligned}\right\} \quad
\begin{aligned}
x\in\mathbb{R}^{n}, u\in\mathbb{R}^{m}, \\
p\in\mathcal{P}, \tau\in[0, \delta],
\end{aligned}  \\
\label{eqn-37}
\left.\begin{aligned}
x^{+}&=D_{p}x \\
p^{+}&\in\mathcal{P}\\
\tau^{+}&=0
\end{aligned}\right\} \quad
\begin{aligned}
x\in\mathbb{R}^{n}, u\in\mathbb{R}^{m}, \\
p\in\mathcal{P}, \tau=\delta,
\end{aligned}
\end{align}
where $\hat{x}\in\mathbb{R}^{n}$ is the delayed state trajectory, $p\in\mathcal{P}$ is the switching signal taking value in finite set $\mathcal{P}$, $\tau$ is a timer to track impulsive switching intervals with bound $\delta>0$, and $A_{p}, B_{p}, C_{p}, D_{p}$ are matrices with appropriate dimensions. The system \eqref{eqn-36}-\eqref{eqn-37} corresponds to a system $\mathcal{H}^{\Delta}_{\mathcal{M}}=(\mathcal{F}, \mathcal{G}, \mathcal{C}, \mathcal{D}, \mathbb{R}^{n}, \mathbb{R}^{m})$:
\begin{align*}
\mathcal{C}&:=\{\varphi=(\psi, p, \tau)\in\mathcal{M}^{\Delta}|\tau(0, 0)\in[0, \delta], p(0, 0)\in\mathcal{P}\}, \\
\mathcal{F}(\varphi, u)&:=\begin{bmatrix}  A_{p}\psi(0, 0)+B_{p}\psi(-r, k(r))+C_{p}u(0, 0) \\ 0 \\ 1 \end{bmatrix}, \\
\mathcal{D}&:=\{\varphi=(\psi, p, \tau)\in\mathcal{M}^{\Delta}|\tau(0, 0)=\delta, p(0, 0)\in\mathcal{P}\}, \\
\mathcal{G}(\varphi, u)&\in\begin{bmatrix} D_{p}\psi(0, 0)   \\ \mathcal{P} \\  0 \end{bmatrix},
\end{align*}
where $k(s)=\max\{k|(s, k)\in\dom\varphi\}$. Note that the system $\mathcal{H}^{\Delta}_{\mathcal{M}}$ corresponds to the switched delayed system $\dot{x}(t)=A_{p}x(t)+B_{p}x(t-r)+C_{p}u(t)$ with the impulse $x^{+}=D_{p}x$ every $\delta$ unit of time.

Let $\mathcal{W }:=\{0\}\times\mathcal{P}\times[0, \delta]$. Since $\varphi= (\psi, p, \tau)$ and $\tau\in[0, \delta]$, we have, by the definitions of $|\cdot|_{\mathcal{W}}$ and $\|\cdot\|_{\mathcal{W}}$, that $|\varphi(0, 0)|_{\mathcal{W}}=|\psi(0, 0)|$ and $\|\varphi\|_{\mathcal{W}}=\|\psi\|$ for all $\varphi\in\mathcal{C}\cup\mathcal{D}\subset\mathcal{M}^{\Delta}$.

For all $\varphi\in\mathcal{M}^{\Delta}$, define an LKF as follows:
\begin{align*}
V (\varphi):=\sigma_{p}|\psi(0, 0)|^{2}+\mu_{p}\int^{0}_{-r}e^{-\eta\tau(s, k(s))}|\psi(s, k(s))|^{2}ds,
\end{align*}
where $\sigma_{p}, \mu_{p}>0, \eta\geq0$ are constants. It is easy to see that (G.1) is satisfied with $\alpha_{1}(\|\varphi(0, 0)\|_{\mathcal{W}})=\min_{p\in\mathcal{P}}\{\sigma_{p}\}|\psi(0, 0)|^{2}$ and $\alpha_{2}(\|\varphi\|_{\mathcal{W}})=\min_{p\in\mathcal{P}}\{\sigma_{p}+\mu_{p}r\}|\psi(0, 0)|^{2}$.

In the flow, the derivative of $V$ satisfies that, for all $\varphi\in\mathcal{C}$,
\begin{align}
\label{eqn-38}
D^{+}V(\varphi)&\leq2\sigma_{p}\psi^{\top}(0, 0)(A_{p}\psi(0, 0) +B_{p}\psi(-r, k(r))  \nonumber  \\
&\quad +C_{p}u)+\mu_{p}\psi^{\top}(0, 0)\psi(0, 0)  \nonumber  \\
&\quad -\mu_{p}\psi^{\top}(-r, k(-r))\psi(-r, k(-r))  \nonumber \\
&\leq(2\sigma_{p}\lambda_{\max}(A_{p})+\mu_{p}+\sigma_{p}\lambda_{\max}(B_{p})  \nonumber  \\
&\quad +\sigma_{p}\lambda_{\max}(C_{p}))|\psi(0, 0)|^{2} \nonumber  \\
&\quad +(\sigma_{p}\lambda_{\max}(B_{p})-\mu_{p}e^{-\eta\epsilon})|\psi(-r, k(-r))|^{2}  \nonumber  \\
&\quad +\sigma_{p}\lambda_{\max}(C_{p})|u|^{2}.
\end{align}
At the jumps, we have that, for all $\varphi\in\mathcal{D}$,
\begin{equation}
\label{eqn-39}
V(\mathcal{G}(\varphi, u))-V (\varphi)\leq(\lambda^{2}_{\max}(D_{p})-1)|\psi(0, 0)|^{2}.
\end{equation}
Based on \eqref{eqn-38}-\eqref{eqn-39}, consider the following three cases. Define $\Lambda_{p}:=2\sigma_{p}\lambda_{\max}(A_{p})+\mu_{p}+\sigma_{p}\lambda_{\max}(B_{p})+\sigma_{p}\lambda_{\max}(C_{p})$ and $\Omega_{p}:=\sigma_{p}\lambda_{\max}(B_{p})-\mu_{p}e^{-\eta\epsilon}$.

\textbf{Case 1.} For all $p\in\mathcal{P}$, if $\Lambda_{p}<\varpi^{-1}, \Omega_{p}<0, \lambda^{2}_{\max}(D_{p})<1$ and $V(\varphi)\geq\varpi\max_{p\in\mathcal{P}}\{\sigma_{p}\lambda_{\max}(C_{p})\}|u|^{2}$ for some $\varpi>0$, then we have $D^{+}V(\varphi)\leq(\Lambda_{p}-\varpi^{-1})|\psi(0, 0)|^{2}<0$ and $V(g(\varphi, u))-V (\varphi)\leq(\lambda^{2}_{\max}(D_{p})-1)|\psi(0, 0)|^{2}<0$. As a result, by Theorem \ref{thm-4}, $\mathcal{W}$ is ISS for $\mathcal{H}^{\Delta}_{\mathcal{M}}$.

\textbf{Case 2.} For all $p\in\mathcal{P}$, if $-\Lambda_{p}>\Omega_{p}\geq0$ and $\lambda^{2}_{\max}(D_{p})>1$, then the flow dynamics is stable, whereas the jump dynamics is unstable. In this case, it follows from Theorem \ref{thm-6} that if $\ln(\lambda^{2}_{\max}(D_{p})-1)<\lambda$, where $\lambda\in(0, \bar{\lambda})$, $\bar{\lambda}:=\max\{\bar{\lambda}_{p}\}$, and $\bar{\lambda}_{p}$ is the solution to $\bar{\lambda}+\Lambda_{p}+\Omega_{p}e^{\bar{\lambda}(\Delta+1)}=0$, then $\mathcal{W}$ is ISS for $\mathcal{H}^{\Delta}_{\mathcal{M}}$.

\textbf{Case 3.} For all $p\in\mathcal{P}$, if $\Lambda_{p}>0, \Omega_{p}\geq0$ and $\lambda^{2}_{\max}(D_{p})<1$, then the flow dynamics is unstable but the jump dynamics is stable. Therefore, from Theorem \ref{thm-7}, if $\varepsilon\ln(\lambda^{2}_{\max}(D_{p})-1)+(1+\varepsilon)\lambda<0$, where $\varepsilon>0, \lambda\in(0, \bar{\lambda}), \bar{\lambda}:=\max\{\bar{\lambda}_{p}\}$, and $\bar{\lambda}_{p}$ is the solution to $-\bar{\lambda}+\Lambda_{p}+\Omega_{p}e^{\bar{\lambda}(\Delta+1)}=0$, then $\mathcal{W}$ is ISS for $\mathcal{H}^{\Delta}_{\mathcal{M}}$.

\section{Conclusion}
\label{sec-conclusion}

In this paper, we studied input-to-state stability of hybrid systems with memory. Both Razumikhin-type and Krasovskii-type stability conditions were derived for hybrid systems with memory. Furthermore, some extensions and relaxations were presented. Finally, the obtained results were illustrated via two numerical examples. Future directions focus on stochastic hybrid systems with memory.

\bibliographystyle{IEEEtran}

\begin{thebibliography}{10}
\bibitem{Goebel2012hybrid}
R.~Goebel, R.~G. Sanfelice, and A.~R. Teel, \emph{Hybrid Dynamical Systems:
  Modeling, Stability, and Robustness}.\hskip 1em plus 0.5em minus 0.4em\relax
  Princeton University Press, 2012.

\bibitem{Cai2007smooth}
C.~Cai, A.~R. Teel, and R.~Goebel, ``Smooth {L}yapunov functions for hybrid
  systems--{P}art {I}: existence is equivalent to robustness,'' \emph{IEEE
  Transactions on Automatic Control}, vol.~52, no.~7, pp. 1264--1277, 2007.

\bibitem{Ren2016stability}
W.~Ren and J.~Xiong, ``Stability and stabilization of switched stochastic
  systems under asynchronous switching,'' \emph{Systems \& Control Letters},
  vol.~97, pp. 184--192, 2016.

\bibitem{Wu2016input}
X.~Wu, Y.~Tang, and W.~Zhang, ``Input-to-state stability of impulsive
  stochastic delayed systems under linear assumptions,'' \emph{Automatica},
  vol.~66, pp. 195--204, 2016.

\bibitem{Liu2011input}
J.~Liu, X.~Liu, and W.-C. Xie, ``Input-to-state stability of impulsive and
  switching hybrid systems with time-delay,'' \emph{Automatica}, vol.~47,
  no.~5, pp. 899--908, 2011.

\bibitem{Ren2017stability}
W.~Ren and J.~Xiong, ``Stability analysis of impulsive stochastic nonlinear
  systems,'' \emph{IEEE Transactions on Automatic Control}, vol.~62, no.~9, pp.
  4791--4797, 2017.

\bibitem{Cai2009characterizations}
C.~Cai and A.~R. Teel, ``Characterizations of input-to-state stability for
  hybrid systems,'' \emph{Systems \& Control Letters}, vol.~58, no.~1, pp.
  47--53, 2009.

\bibitem{Goebel2006solutions}
R.~Goebel and A.~R. Teel, ``Solutions to hybrid inclusions via set and
  graphical convergence with stability theory applications,''
  \emph{Automatica}, vol.~42, no.~4, pp. 573--587, 2006.

\bibitem{Medvedeva2015synthesis}
I.~V. Medvedeva and A.~P. Zhabko, ``Synthesis of {R}azumikhin and
  {L}yapunov--{K}rasovskii approaches to stability analysis of time-delay
  systems,'' \emph{Automatica}, vol.~51, pp. 372--377, 2015.

\bibitem{Zhou2016razumikhin}
B.~Zhou and A.~V. Egorov, ``Razumikhin and {K}rasovskii stability theorems for
  time-varying time-delay systems,'' \emph{Automatica}, vol.~71, pp. 281--291,
  2016.

\bibitem{Heemels2010networked}
W.~M.~H. Heemels, A.~R. Teel, N.~Van~de Wouw, and D.~Ne{\v{s}}i{\'c},
  ``Networked control systems with communication constraints: {T}radeoffs
  between transmission intervals, delays and performance,'' \emph{IEEE
  Transactions on Automatic control}, vol.~55, no.~8, pp. 1781--1796, 2010.

\bibitem{Liu2016lyapunov}
J.~Liu and A.~R. Teel, ``Lyapunov-based sufficient conditions for stability of
  hybrid systems with memory,'' \emph{IEEE Transactions on Automatic Control},
  vol.~61, no.~4, pp. 1057--1062, 2016.

\bibitem{Liu2014hybrid}
------, ``Hybrid systems with memory: modelling and stability analysis via
  generalized solutions,'' \emph{Proceedings of IFAC}, pp. 6019--6024, 2014.

\bibitem{Naghshtabrizi2010stability}
P.~Naghshtabrizi, J.~P. Hespanha, and A.~R. Teel, ``Stability of delay
  impulsive systems with application to networked control systems,''
  \emph{Transactions of the Institute of Measurement and Control}, vol.~32,
  no.~5, pp. 511--528, 2010.

\bibitem{Yuan2016delay}
C.~Yuan and F.~Wu, ``Delay scheduled impulsive control for networked control
  systems,'' \emph{IEEE Transactions on Control of Network Systems}, 2016.

\bibitem{Khadra2005impulsively}
A.~Khadra, X.~Liu, and X.~Shen, ``Impulsively synchronizing chaotic systems
  with delay and applications to secure communication,'' \emph{Automatica},
  vol.~41, no.~9, pp. 1491--1502, 2005.

\bibitem{Zhu2014mean}
Q.~Zhu and J.~Cao, ``Mean-square exponential input-to-state stability of
  stochastic delayed neural networks,'' \emph{Neurocomputing}, vol. 131, pp.
  157--163, 2014.

\bibitem{Chen2009input}
W.-H. Chen and W.~X. Zheng, ``Input-to-state stability and integral
  input-to-state stability of nonlinear impulsive systems with delays,''
  \emph{Automatica}, vol.~45, no.~6, pp. 1481--1488, 2009.

\bibitem{Dashkovskiy2012stability}
S.~Dashkovskiy, M.~Kosmykov, A.~Mironchenko, and L.~Naujok, ``Stability of
  interconnected impulsive systems with and without time delays, using
  {L}yapunov methods,'' \emph{Nonlinear Analysis: Hybrid Systems}, vol.~6,
  no.~3, pp. 899--915, 2012.

\bibitem{Liu2006stability}
X.~Liu and J.~Shen, ``Stability theory of hybrid dynamical systems with time
  delay,'' \emph{IEEE transactions on automatic control}, vol.~51, no.~4, pp.
  620--625, 2006.

\bibitem{Liu2018hybrid}
J.~Liu and A.~R. Teel, ``Hybrid systems with memory: existence and
  well-posedness of generalized solutions,'' \emph{SIAM Journal on Control and
  Optimization}, vol.~56, no.~2, pp. 1011--1037, 2018.

\bibitem{Sontag1995characterizations}
E.~D. Sontag and Y.~Wang, ``On characterizations of the input-to-state
  stability property,'' \emph{Systems \& Control Letters}, vol.~24, no.~5, pp.
  351--359, 1995.

\bibitem{Jiang2001input}
Z.-P. Jiang and Y.~Wang, ``Input-to-state stability for discrete-time nonlinear
  systems,'' \emph{Automatica}, vol.~37, no.~6, pp. 857--869, 2001.

\bibitem{Liu2016razumikhin}
K.-Z. Liu and X.-M. Sun, ``Razumikhin-type theorems for hybrid system with
  memory,'' \emph{Automatica}, vol.~71, pp. 72--77, 2016.

\bibitem{Sanfelice2014input}
R.~G. Sanfelice, ``Input-output-to-state stability tools for hybrid systems and
  their interconnections,'' \emph{IEEE Transactions on Automatic Control},
  vol.~59, no.~5, pp. 1360--1366, 2014.

\bibitem{Teel1998connections}
A.~R. Teel, ``Connections between {R}azumikhin-type theorems and the {ISS}
  nonlinear small gain theorem,'' \emph{IEEE Transactions on Automatic
  Control}, vol.~43, no.~7, pp. 960--964, 1998.

\bibitem{Sontag1996new}
E.~D. Sontag and Y.~Wang, ``New characterizations of input-to-state
  stability,'' \emph{IEEE Transactions on Automatic Control}, vol.~41, no.~9,
  pp. 1283--1294, 1996.

\end{thebibliography}

\end{document}